\documentclass[prc,superscriptaddress,showpacs,showkeys,twocolumn,floatfix]{revtex4}

\usepackage{amsmath}
\usepackage{amssymb}
\usepackage{graphicx}
\usepackage{dcolumn}
\usepackage{bm}

\newcommand{\bD}{$b_{D}=(6.6649 \pm 0.0040)$~fm}
\newcommand{\bdmolourvale}{6.6649}
\newcommand{\bdmolourunc}{6.0$\times10^{-4}$} 
\newcommand{\bnd}{$b_{\mathit{nd}}=(6.6649 \pm 0.0040)$~fm}
\newcommand{\bdcorourval}{6.6649}
\newcommand{\bdcorourunc}{6.0$\times10^{-4}$} 
\newcommand{\bndprevworld}{$b_{\mathit{nd}}=(6.6727\pm0.0045)$~fm}
\newcommand{\bdprevworldval}{6.6727}
\newcommand{\bdprevworldunc}{6.7$\times10^{-4}$} 
\newcommand{\bndcurworld}{$b_{\mathit{nd}}=(6.6683\pm0.0030)$~fm}
\newcommand{\bdcurworldval}{6.6683}
\newcommand{\bdcurworldunc}{4.5$\times10^{-4}$} 
\newcommand{\DoubletExpPrev}{$^{2}a_{\mathit{nd}}=(0.65\pm0.04$(expt))~fm}
\newcommand{\DoubletExpTheoryNew}{$^{2}a_{\mathit{nd}}=(0.645\pm0.003$(expt)~$\pm~0.007$(theory))~fm}
\newcommand{\QuartetTheory}{$^{4}a_{\mathit{nd}}=(6.346\pm0.007)$~fm}

\newcommand{\bH}{$b_{H}=(-3.7458 \pm 0.0020)$~fm}
\newcommand{\bpmolourvalue}{-3.7458}
\newcommand{\bpmolourunc}{5.3$\times10^{-4}$} 
\newcommand{\bnp}{$b_{\mathit{np}}=(-3.7384 \pm 0.0020)$~fm}
\newcommand{\bpcorourval}{-3.7384}
\newcommand{\bpcorourunc}{5.3$\times10^{-4}$} 
\newcommand{\bpprevworldval}{-3.7410}
\newcommand{\bpcurworldval}{-3.7405}
\newcommand{\NowakSymbol}{$\Delta b_{Nowak}$}
\newcommand{\NowakCorrectionH}{$\Delta b_{Nowak}=1.9\times10^{-3}\times(-3.74~$fm$)=-0.007$~fm}
\newcommand{\OurExpDifference}{$-0.0048$~fm}
\newcommand{\bne}{$b_{ne}=(-1.33\pm0.03)\times10^{-3}$~fm}
\newcommand{\bcarbon}{$b_{C}=(6.6484\pm0.0013)$~fm}

\newcommand{\bnpprevworld}{$b_{\mathit{np}}=(-3.7410\pm0.0010)$~fm}
\newcommand{\bnpcurworld}{$b_{\mathit{np}}=(-3.7405\pm0.0009)$~fm}

\newcommand{\bpprevworldunc}{2.7$\times10^{-4}$} 
\newcommand{\bpcurworldunc}{2.5$\times10^{-4}$} 

\newcommand{\etal}{\textit{et al.}}

\begin{document}
\thepage
\title{Precision neutron interferometric measurements \\
and updated evaluations of the n-p and n-d coherent neutron
scattering lengths}

\author{K. Schoen}
\affiliation{University of Missouri-Columbia, Columbia, MO  65211,
USA}
\author{D.\,L. Jacobson}
\affiliation{National Institute of Standards and Technology,
Gaithersburg, MD 20899-8461, USA}
\author{M. Arif}
\affiliation{National Institute of Standards and Technology,
Gaithersburg, MD 20899-8461, USA}
\author{P.\,R. Huffman}
\affiliation{National Institute of Standards and Technology,
Gaithersburg, MD 20899-8461, USA}
\author{T.\,C. Black}
\affiliation{University of North Carolina at Wilmington,
Wilmington, NC 28403-3297, USA}
\author{W.\,M. Snow}
\affiliation{Indiana University/IUCF, Bloomington, IN 47408, USA}
\author{S.\,K. Lamoreaux}
\affiliation{Los Alamos National Laboratory, Los Alamos, NM 87545,
USA }
\author{H. Kaiser}
\affiliation{University of Missouri-Columbia, Columbia, MO  65211,
USA}
\author{S.\,A. Werner}
\affiliation{University of Missouri-Columbia, Columbia, MO  65211,
USA} \affiliation{National Institute of Standards and Technology,
Gaithersburg, MD 20899-8461, USA}

\date{\today}

\begin{abstract}
We have performed high precision measurements of the coherent neutron
scattering lengths of gas phase molecular hydrogen and deuterium using
neutron interferometry.  After correcting for molecular binding and
multiple scattering from the molecule, we find \bnp\ and \bnd.  Our
results are in agreement with the world average of previous
measurements, \bnpprevworld\ and \bndprevworld.  The new world
averages for the n-p and n-d coherent scattering lengths, including
our new results, are \bnpcurworld\ and \bndcurworld.  We compare
$b_{\mathit{nd}}$ with the calculations of the doublet and quartet
scattering lengths of several nucleon-nucleon (NN) potential models
and show that almost all known calculations are in disagreement with
the precisely-measured linear combination corresponding to the
coherent scattering length.  Combining the world data on
$b_{\mathit{nd}}$ with the modern high-precision theoretical
calculations of the quartet n-d scattering lengths recently summarized
by Friar \etal, we deduce a new value for the doublet scattering
length of
$^{2}a_{\mathit{nd}}=(0.645\pm0.003$(exp)$\pm0.007$(theory))~fm.  This
value is a factor of 4 more precise than the previously accepted value
of $^{2}a_{\mathit{nd}}=(0.65\pm0.04$(exp))~fm.  The current state of
knowledge of scattering lengths in the related p-d system, ideas for
improving by a factor of five the accuracy of the $b_{\mathit{np}}$
and $b_{\mathit{nd}}$ measurements using neutron interferometry, and
possibilities for further improvement of our knowledge of the coherent
neutron scattering lengths of $^{3}$H, $^{3}$He, and $^{4}$He are
discussed.
\end{abstract}


\maketitle

\section{Introduction}
\label{sect:intro}

The three-nucleon system is both fascinating and remarkable.  Despite
decades of intensive study with increasingly sophisticated theoretical
tools, fundamental facets of the system remain mysterious, confusing,
and contradictory.  Realistic nucleon-nucleon (NN) potentials
underbind $^3$He and $^3$H by several hundred keV \cite{Fri83}.  The
p-d and n-d scattering lengths in the doublet \textit{s}-wave channel,
which are apparently strongly correlated to the $^3$He and $^3$H
binding energies respectively, are likewise poorly predicted by NN
force models.  Convergence between theory and experiment for these
fundamental parameters can be obtained only by the \textit{ad hoc}
admixture of three-nucleon (3N) forces.  Although it is well
understood that 3N forces must exist with a weaker strength and
shorter range than the NN force, little else is known.  The incomplete
nature of current 3N force models is demonstrated by their tendency to
resolve certain problems, such as tri-nucleon underbinding, at the
expense of exacerbating discrepancies in other observables
\cite{Ish99,Wit93,Glo96,Kie95a,Ros95,Kie95b,Pud97,Car94}.

Of course one must first understand two-nucleon forces in detail
before one can isolate possible 3N force effects.  A wide variety of
NN forces are deployed in modern nuclear physics calculations, but
they have certain features in common.  They all employ short-range,
semi-phenomenological forces matched to a one pion exchange potential
(OPEP) tail beyond $\approx 1.4$~fm, and they all give roughly the
same results for the low-energy observables.  They are all constrained
to reproduce the NN observables as well as the properties of the
deuteron.  They also must incorporate deviations from isospin symmetry
to describe the data.  One of the most sophisticated of the modern
potentials--the AV18 potential--includes, in addition to purely
electromagnetic terms, terms accounting for charge independence
breaking (CIB) and charge symmetry breaking (CSB), which are
phenomenologically adjusted to replicate the n--n, p--p and n--p
scattering lengths\cite{Wir95}.

The accessible places to look for 3N force effects are the bound
states of $^{3}$H and $^{3}$He and the scattering states of n-d and
p-d.  Since $^{3}$H and n-d are free from electromagnetic
complications (both theoretical and experimental), they are the
systems of choice for precision tests.  For all of these systems, the
computational tools presently available are believed to be excellent. 
``Exact'' Fadeev solutions are available for the scattering states
below the deuteron breakup threshold.  Above breakup threshold, the
n-d equations can also be solved exactly throughout the entire range
of applicability of the potential models employed.  In addition to
these exact models, other methods based on the reactance matrix
(\textit{K}-matrix), such as the hyperspherical harmonic expansion of
Kievsky \etal\cite{Kie93}, the effective field theory of Bedaque
\etal\cite{Bed97}, as well as dispersion theory/phenomenological
models such as that of Hale\cite{Hal00}, provide complementary insights
and information that goes beyond tests of potential models.

Recently, significant insight into certain features of three-nucleon
systems has come from the effective field theory (EFT) approach
\cite{Bed98, Ham99}.  EFT has been used to solve the three nucleon
problem with short-range interactions in a systematic expansion in the
small momentum region set by $kb \leq 1$, where $k$ is the momentum
transfer and $b$ is the scattering length.  For the two-body system
EFT is equivalent to effective range theory and reproduces its
well-known results \cite{Kol98,Kap98,Geg98}.  The chiral EFT expansion
does not require the introduction of an operator corresponding to a 3N
force until next-to-next-to leading order (NNLO) in the expansion, and
at this order it requires only two low energy constants\cite{Epe01}. 
With these two parameters determined from experiment, chiral EFT can
make precise predictions for other three-body observables.  It is
clear that precision measurements of the three-nucleon scattering
lengths provide both valuable theoretical benchmarks and critical
input parameters for the study of three-nucleon systems.

The most important data, of course, is the measurement of the
zero-energy scattering.  With the neutron interferometer at NIST, we
have made measurements of the coherent n-p and n-d bound scattering
lengths, $b$, which are related to the free scattering length $a$ by
\begin{equation}
    \label{eq:relofbtoa}
    a = \frac{M}{m+M}b.
\end{equation}
Here, $m$ is the mass of the neutron and $M$ is the mass of the atom. 
For hydrogen, $a$ is the linear combination of the singlet and the
triplet scattering lengths given by,
\begin{equation}
    \label{eq:compnp}
    a_{\mathit{np}} = (1/4)~^{1}a_{\mathit{np}} + 
			(3/4)~^{3}a_{\mathit{np}},
\end{equation} and for deuterium it is the linear combination of the
    doublet and quartet scattering lengths,
\begin{equation}
    \label{eq:compnd}
    a_{\mathit{nd}} = (1/3)~^{2}a_{\mathit{nd}} + 
			(2/3)~^{4}a_{\mathit{nd}}.
\end{equation}
These new measurements have uncertainties that are comparable to (in
n-p) and smaller than (in n-d) previous results.  

The n--d measurement yields a linear combination of the doublet and
quartet scattering lengths and not the separate channel scattering
lengths.  There are good reasons however to have confidence that the
doublet ($^2$S$_{1/2}$) scattering length can be extracted from this
measurement reliably with some help from theory.  In the first place,
the quartet $s$-wave scattering length ($^4$S$_{3/2}$) can be
unambiguously determined.  Because the three nucleons in this channel
exist in a spin-symmetric state, and hence have an antisymmetric
space-isospin wavefunction, the scattering in this state is completely
determined by the long range part of the triplet $s$-wave NN
interaction in the n-p channel; i.e. by n-p scattering and the
properties of the deuteron.  Furthermore, the use of multi-energy
phase shift analysis, such as that conducted by Black \etal\ in the
p-d system\cite{Bla99}, could allow one to extrapolate to the very low
energy quartet phase shifts on the basis of data at higher energies. 
This is possible because, unlike the $^2$S$_{1/2}$ effective range
function, the $^4$S$_{3/2}$ effective range function is free of
singularities.  Therefore, one can independently predict the zero
energy quartet $s$-wave scattering length $^{4}a_{\rm nd}$ both from
fundamentally sound theory and from the totality of the sub-breakup
n-d database.

A high accuracy determination of the $^2$S$_{1/2}$ scattering length
in the n-d system is of interest for a number of reasons.  The Pauli
principle does not deter this channel from exploring the shorter-range
components of the interaction, where 3N forces should appear, and the
Coulomb interaction is not present to complicate experiment or
analysis.  The n-d system is also in principle sensitive to charge
symmetry breaking (CSB) effects.  Due to charge symmetry breaking, the
n-p force and the n-n force are different.  CSB is explicitly included
in modern NN potentials such as the Argonne AV18 potential.  The
accuracy in this observable that can be achieved in modern
calculations is sufficient to be sensitive to CSB.

It was discovered in the early days of work on nuclear three-body
systems that there is an empirical relation between the calculated
doublet n-d scattering length and the binding energy of the
tri-nucleon system.  These so-called ``Phillips lines'' \cite{Phi77}
have been shown to be strictly linear for the n-p system, (in which
case it is the triplet scattering length that is linked with the
deuteron binding energy) and they are approximately linear for n-d
scattering.  When Phillips lines are plotted for the n-d calculations,
the physical triton binding energy (8.48~MeV) intersects the currently
accepted doublet value $^{2}a_{\mathit{nd}}=0.65$~fm \cite{Che91}, in
excellent agreement with the measurement of Dilg \etal\ of
$(0.65\pm0.04)$~fm \cite{Dil71}.  By contrast, a charge-symmetric
R-matrix analysis of n-d data conducted by Hale \cite{Hal00} at
energies up to the three-body breakup threshold gives the preliminary
value $^{2}a_{\mathit{nd}} = 0.41$~fm, even though the Dilg \etal\
measurements were the only ones included in the n-d data set.  These
results, although preliminary, call into question the experimental
values of the n-d scattering lengths, particularly in the doublet spin
state.  If the value of $^{2}a_{\mathit{nd}}$ really is significantly
less than 0.65~fm, then the presumably well-understood phenomenology
of the Phillips line is thrown into disarray.

The current values of the n-d scattering lengths were determined by
measurements performed almost 30~years ago using a neutron gravity
reflectometer \cite{Dil71}.  These measurements were not performed
using pure D$_2$ gas samples because the experimental technique
required samples in liquid or solid form: they were performed on
D$_2$O, SiO$_{2}$ and Si from which the scattering length for n-d was
deduced.  This report describes the first high-precision measurement
of the n-d coherent scattering length performed using a pure sample. 
In addition, we chose to pursue a precision measurement of the n-p
coherent scattering length.  The measurement of $b_{\mathit{np}}$
should yield a result that is consistent with the world average of
other n-p measurements.  If it doesn't then we may have an unaccounted
for systematic effect in our measurement.  Therefore the measurement
of $b_{\mathit{np}}$ provides a secondary check on possible unknown
systematic errors of our method.

We use neutron interferometry to measure the coherent scattering
lengths of these gases.  Interferometry measures a phase shift which
is simply related to the coherent scattering length of the nuclei and
is therefore sensitive to a particular linear combination of the
scattering amplitudes in the two channels.  From an experimental point
of view, it is possible to measure this linear combination to much
higher precision than one can determine the individual scattering
lengths.  For example, the coherent scattering length of silicon has
been measured using neutron interferometry to the incredible accuracy
of $0.005$~\%, surely one of the most accurate measurements of a
nuclear scattering amplitude ever performed\cite{Iof98}.  The
experiment requires neither polarized beams nor polarized targets and
is sensitive to the amplitude as opposed to the square of the
amplitude that enters into cross section measurements.

The remainder of this paper is organized as follows. 
Section~\ref{sect:NeutOptTheory} outlines results from the theory of
neutron optics that are relevant to the interpretation of the
measurements.  Sections~\ref{sect:NeutronInterferometry} and
\ref{sect:PhaseShiftMeas} describe the neutron interferometry
technique and measurement.  Section~\ref{sect:GasCellDesign} covers
the gas cell design, which is also chosen to minimize the sample cell
phase shift.  Section~\ref{sect:GasCellAlignment} describes the
procedure by which the target cell was aligned on the interferometer
to make the measurement insensitive to the large phase shifts
introduced by the gas sample cell.  Section~\ref{sect:DataCollection}
presents the phase shift data, uncorrected at this point for various
slow changes in gas density.  Section~\ref{sect:AtomDensity} describes
the measurements which determine the gas density using the ideal gas
law with virial coefficient corrections.  Section~\ref{sect:GasPurity}
shows the measurements which establish the gas composition of the
deuterium sample, which has non-negligible HD and H$_{2}$
contaminants.  Section~\ref{sect:Thickness} describes the sample
thickness measurements with and without pressurized gas in the cell
and presents the last time-dependent correction to the gas density,
which removes all time dependence to the phase shift data. 
Section~\ref{sect:Wavelength} outlines the measurement of the neutron
wavelength and describes the measurement that establishes its
stability over time.  Section~\ref{sect:ExperimentalResults} relates
our measurement of the coherent scattering lengths of H$_{2}$ and
D$_{2}$ molecules to the nuclear scattering lengths, outlines the
correction to neutron optical theory due to the breakdown of the
impulse approximation, presents our final values for the scattering
lengths, and compares them with past measurements. 
Section~\ref{sect:CompWithTheory} describes the comparison of the
world's data on the n-d coherent scattering length to existing
theoretical calculations and uses the recently-available high
precision calculations of the quartet n-d scattering length in second
generation NN potential models to determine the doublet n-d scattering
length.  It also places the measurements in the n-d system into the
wider context of p-d measurements.  Finally section
~\ref{sect:Conclusion} summarizes our results and their significance
and discusses the possibilities for further improvements in the
precision of coherent n-p and n-d scattering length measurements as
well as the possibilities for high-precision coherent scattering
length measurements in n-$^{3}$He, n-$^{4}$He, and n-$^{3}$H systems. 
Appendix~\ref{sect:NowakCorrection} describes in more detail the small
corrections to the theory derived by Nowak, to which our experiment is
sensitive.  Appendix~\ref{sect:WorldAverageCalculation} discusses the
procedure used to determine the world average using published values
of the scattering lengths.

\section{Neutron Optics Theory}
\label{sect:NeutOptTheory}
In this section we give a brief review of the relevant results
from the theory of neutron optics that are needed to understand
precisely how the phase shift measured in the interferometer is
related to the neutron scattering lengths of interest. For a more
detailed treatment see Sears\cite{Searsbook}.

Neutron optics is based on the existence of the ``coherent wave''
which is the coordinate representation of the coherent state
formed  by the incident wave and the forward scattered wave in a
scattering  medium. It is determined by the solution of a one-body
Schrodinger equation
\begin{equation}
    \label{eq:coherent}
    \left[\frac{-\hbar^{2}}{ 2m}\Delta +v(r)\right]\psi(r)=E\psi(r)
\end{equation}
where $\psi(r)$ is the coherent wave and $v(r)$ is the optical
potential of the medium. The coherent wave satisfies the
Lippmann-Schwinger equation
\begin{equation}
    \label{eq:lippmann}
    \psi(r)=\left|k\right>+gv(r)\psi(r)
\end{equation}
where $\left|k\right>$ is the incident wave, $g$ is the one-body
Green's function for nonrelativistic motion  of a neutron and
$v(r)$ is the optical potential. The optical potential is related
to the one body $t$ matrix by
\begin{equation}
    \label{eq:tmatrix}
    t=v(r)+tgv(r)
\end{equation}
and this combination forms the usual coupled system of equations
of nonrelativistic scattering theory. Given a form for the $t$
matrix  one can determine the optical potential and then solve the
one-body Schrodinger equation for the coherent wave.

One must make an approximation for the $t$ matrix of the neutron
in a medium of scatterers. The usual approximation is essentially
the Born approximation in which $v=t$ and
\begin{equation}
    \label{eq:tapprox}
    t=\sum_{l}t_{l}.
\end{equation}
Finally one must approximate the one-body $t$ matrix $t_{l}$.
Using the impulse approximation for scattering, one gets
\begin{equation}
    \label{eq:FermiApprox}
    t_{l}=~(2\pi \hbar^2/m)\sum_{l} b_{l} \delta({\bf r}-{\bf R}_{l}).
\end{equation}
Here, $l$ denotes the elemental species, $b_{l}$ is the coherent
scattering length for element $l$, ${\bf r}$ is a random spatial
coordinate and ${\bf R}_{l}$ defines the coordinate of each atom
the neutron can scatter from. From Eq.~\ref{eq:FermiApprox} we
then arrive at an expression for the optical potential
\begin{equation}
    \label{eq:opticalpotential}
    v_{opt}(r)~=~(2\pi \hbar^2/m)\sum_{l} N_{l} b_{l},
\end{equation}
where $N_{l}$ is the number density of scatterers.

The effect of the optical potential on the beam for a nonabsorbing
uniform medium is to slow down the neutrons  as they encounter the
potential step due to the matter, thereby decreasing the neutron
wave vector, \textit{K}, within the medium. The neutron index of
refraction is defined by this relative change in the magnitude of
the wavevector $n=K/k$. Conservation of energy at the boundary
determines the relation to the optical potential
\begin{equation}
    \label{eq:refractionindex}
    n^{2}=1-{v_{0}/E}.
\end{equation}

In general the scattering amplitude is complex to account for
incoherent scattering and absorption of the wave amplitude.  The
imaginary part of the scattering amplitude is related to the total
reaction cross section by the optical theorem and leads to the
more accurate form of the index of refraction
\begin{eqnarray}
 \label{eq:optindex}
    n & = & n_{r} + i n_{i} \nonumber \\
    &\approx & 1 - \sum_{l} \biggl[ \left( N_{l}
    \lambda^{2} / 2 \pi\right) \sqrt{b^{2} - \left(\sigma_{l} /2
    \lambda\right)^{2}} \biggr. \\
    & & \mbox{} + \biggl.i N_{l} \sigma_{l} \left(\lambda / 4
    \pi\right) \biggr]. \nonumber
\end{eqnarray}
Here, $n_{r}$ ($n_{i}$) is the real (imaginary) part of the index
of refraction, $\lambda$ is the neutron deBroglie wavelength, $b$
is the bound coherent scattering length, and $\sigma _{l}$ is the
total reaction cross section (scattering plus absorption). For
typical neutron-nucleus potentials, the real part of $n$ is near
unity, $(n_{r}- 1)\approx10^{-5}$.  In less accurate treatments,
the second term in the square root of the real part is omitted.
However, its presence is required by the optical theorem and is
included here for completeness.

In neutron optics experiments with unpolarized neutrons and
unpolarized samples the coherent scattering length $b$ is the sum
of the scattering lengths in both scattering channels weighted by
the number of spin states in each channel.  From a quantum
mechanical point of view, this is the total amplitude for a
neutron to scatter without a change in the internal state of the
target.  In the forward direction only this amplitude can
interfere with the unscattered incident wave. In atomic deuterium
with nuclear spin $S=1$, for example, the coherent scattering
length is the weighted sum of the doublet ($S=1/2$) and quartet
($S=3/2$) states
\begin{equation}
    \label{eq:b_deuterium}
    b_{\mathit{nd}} = \frac{M+m}{M} \left[(1/3)\,^{2}a_{\mathit{nd}}
                    +              (2/3)\,^{4}a_{\mathit{nd}}\right],
\end{equation}
and for atomic hydrogen it is the weighted sum of the
corresponding singlet  and triplet states
\begin{equation}
    \label{eq:b_hydrogen}
    b_{\mathit{np}} = \frac{M+m}{M}\left[(1/4)\,^{1}a_{\mathit{np}}
                    +              (3/4)\,^{3}a_{\mathit{np}}\right].
\end{equation}
Here the superscripts label the number of spin states for each
scattering channel. In the next section the relation between the
phase shift of the coherent wave in the interferometer and the
bound coherent scattering length is derived.

\section{Experimental Procedure and Results}
\subsection{Apparatus and basic ideas of neutron interferometry}
\label{sect:NeutronInterferometry}
\begin{figure}[t]
    \begin{center}
    \includegraphics[width=3.228in]{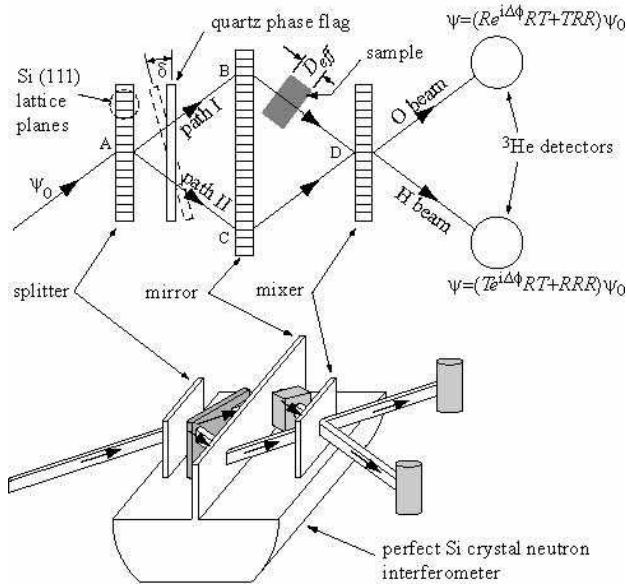}
    \end{center}
    \caption{ A schematic view of the Si perfect crystal neutron
    interferometer.  Parameters associated with the neutron optics are
    discussed in the text.}
    \label{fig:Interferometer}
\end{figure}

Scattering length measurements were performed at the National
Institute of Standards and Technology (NIST) Center for Neutron
Research (NCNR) Interferometer and Optics Facility \cite{Ari94}.  This
facility, situated on cold neutron guide seven (NG7), consists of a
perfect crystal silicon neutron interferometer with high phase
contrast ($80~\%$) and long-term phase stability ($\approx5~^{\circ}$C
per day).  The single crystal interferometer is schematically
illustrated in Fig.~\ref{fig:Interferometer}.  A monochromatic cold
neutron beam ($E = 11.1$~MeV, $\lambda = 0.271$~nm, $\Delta \lambda /
\lambda \leq 0.5~\%$) is diffracted into the facility using a pair of
pyrolytic graphite (PG) crystals operated in a non-dispersive double
crystal mode using the (002) reflection.  This beam is collimated and
vertically focused onto the interferometer.  The neutron beam is
coherently divided in the splitter slab near point A by Bragg
diffraction into two beams that travel along paths $I$ and $II$. 
These beams are again split in the mirror slab near points B and C by
Bragg diffraction.  Two of the wave fields are coherently recombined
to interfere in the third Si mixer slab near point D. The perfect
single crystal nature of the device ensures the required alignment
precision which allows the Bragg diffraction condition to be met by
all three silicon slabs simultaneously.  In addition, the narrow phase
space acceptance of the device is an efficient filter for all
incoherent neutron interactions in the silicon, phase shifter, and
samples.

The geometry traced out by this kind of interferometer is commonly
referred to as Mach-Zehnder by analogy with similar optical
interferometers.  Here, however, the splitter, mirror and mixer
crystals have been machined from a monolithic Si ingot in which the
lower backbone maintains the perfect lattice registry of the original
ingot.  This avoids the mechanical difficulties associated with
permanently aligning separated crystals to tolerances much less than
the angular acceptance of a perfect crystal ($5~\mu$rad).  The
interferometer crystal geometry is denoted as a Laue-Laue-Laue (LLL)
crystal since there are three transmission Bragg reflecting crystals
in the Laue geometry.  For a detailed description of neutron
interferometry techniques and experiments, refer to the book by Rauch
and Werner \cite{Werbook}.

The relative intensities recorded for the two exit beams, called the
O-beam and the H-beam here, depend upon the phase difference of the
neutron waves traversing path $II$ relative to path $I$.  The phase
shift that we measure is due to the phase difference of the neutron
wave traversing path $I$ with the sample in this path relative to its
phase with the sample removed.  For the samples used in these
measurements, the attenuation cross section does not significantly
influence the real part of the refractive index.  To very high
accuracy, the phase shift due to the sample can be written as,
\begin{equation}
    \label{eq:phase}
    \Delta \phi = (n_{r} - 1) k D_{\mathit{eff}} = - \sum_{l} \lambda
N_{l}
    b_{l} D_{\mathit{eff}},
\end{equation}
where $D_{\mathit{eff}}$ is the effective neutron optical thickness of
the sample medium along the direction of wave propagation.  Thus, a
measurement of the coherent scattering length $b$ to $0.05$~\%
absolute accuracy demands an absolute measurement of the following
quantities at the $0.02$~\% level:

(1) the neutron optical phase shift of the gas, $\Delta \phi$,

(2) the atom density $N$,

(3) the sample thickness $D_{\mathit{eff}}$, and

(4) the wavelength $\lambda$. 

\noindent In addition, it is necessary to verify the purity of the gas
at the same level of accuracy.  Sections \ref{sect:PhaseShiftMeas}
through ~\ref{sect:Wavelength} discuss these measurements in detail.

\subsection{Measurement of the molecular H \& D gas phase shift}
\label{sect:PhaseShiftMeas}
The phase shift, $\Delta\phi$, is measured by a secondary sampling
method in which a flat phase shifter plate (denoted as the quartz
phase flag in Fig.~\ref{fig:Interferometer}) is positioned between
the splitter and mixer blades to intercept both neutron paths.
This phase flag has an optical thickness described by
\begin{subequations}
    \label{eq:Deff} 
    \begin{eqnarray}
     \Delta D^{\mathit{flag}}& = & D^{flag}_{II} - D^{flag}_{I}  =
2D^{flag}_{0}\xi(\Delta \delta),\label{equationa}
    \\
     \xi (\delta) & = & \frac{\sin(\theta_{B}) \sin(\Delta
\delta)}{\cos^{2}(\theta_{B}) - \sin^{2}(\Delta
    \delta)}.\label{equationb}
    \end{eqnarray}
\end{subequations}
Here $D_{0}$ is the thickness of this quartz flag (1.5~mm),
$\theta_{B}$ is the Bragg angle of the neutron beam reflecting
from interferometer Si (111) lattice planes, and
$\Delta\delta=(\delta - \delta_{0})$ is the rotation angle offset
of the phase flag in the horizontal plane.

The intensities of the O and H beams (see
Fig.~\ref{fig:Interferometer}) as a function of the phase flag
angle, $\Delta \delta$, are referred to as interferograms and can
be described by the following relations:
\begin{eqnarray}
    \label{eq:IO}
    I_{O}(\delta) & = & \left| R R T \psi_{0} 
                        + T e^{-i\Delta\phi(\delta)}
                        R R \psi_{0} \right|^{2} \\
                  & = & A_{O} + B \cos\left( C \xi(\delta)
             + \Delta\phi_{0}'\right) \nonumber
\end{eqnarray}
and
\begin{eqnarray}
    \label{eq:IH}
    I_{H}(\delta) & = & \left| T R T \psi_{0} 
         + R e^{-i\Delta\phi(\delta)} R R \psi_{0} \right|^{2} \\
                  & = & A_{H} \nonumber + B \cos\left( C \xi(\delta) 
	 + \Delta\phi_{0}' + \pi \right). \nonumber
\end{eqnarray}
Here $R$ or $T$ is the Si crystal reflection or transmission
coefficient, and the parameters $A_{O}$, $A_{H}$, $B$, $C$, and
$\Delta\phi_{0}'$ are extracted from fits to the data.  The value
of $\Delta\phi_{0}'$ and its corresponding uncertainty is used to
determine the phase difference between the two interfering beams.

\begin{figure}[t]
    \begin{center}
    \includegraphics[width=3.164in]{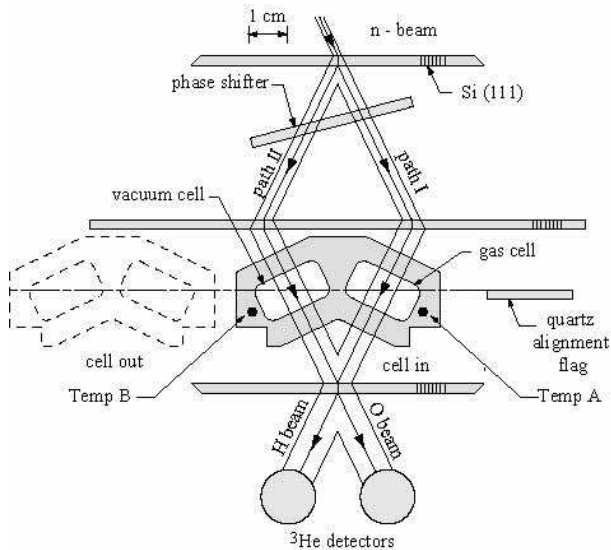}
    \end{center}
    \caption{Schematic view of the interferometer with gas cell and
    quartz alignment flag.  The quartz alignment flag is shown in the
    out position and the centerline denotes an end on view of the
    kinematic mounting plane.  The position of the temperature probes
    (labelled Temp A and Temp B) are shown on the gas cell.}
    \label{fig:InterferometerNGasCell}
\end{figure}

This experiment measures the phase shift when a gas sample is
placed into neutron path $I$ (see
Fig.~\ref{fig:InterferometerNGasCell}).  The gas and sample
housing (cell) contribute separate phase shifts,
$\Delta\phi_{0}'\rightarrow\Delta\phi_{\mathit{gas}}
+\Delta\phi_{\mathit{cell}}+\Delta\phi_{0}$.
The effective thickness of the sample depends upon misalignment
according to,
\begin{equation}
    \label{eq:Deff2}
    D_{\mathit{eff}}(\varepsilon,\gamma) =
(D_{0})/[\cos(\Delta\varepsilon)\cos(\Delta\gamma)],
\end{equation}
where $D_{0}$ is the true thickness and $\Delta\varepsilon =
(\varepsilon - \varepsilon_{0})$ and $\Delta\gamma = (\gamma -
\gamma_{0})$ correspond to the horizontal rotation and vertical
tilt of the sample.  The condition $\varepsilon = \varepsilon_{0}$
and $\gamma=\gamma_{0}$ occurs when the beam is incident normally
on the sample.  In this experiment
$\Delta\phi_{\mathit{gas}}+\Delta\phi_{\mathit{cell}}+\Delta\phi_{0}$
is measured when the beam is incident normally.  In this
orientation the phase shift is insensitive in first order to the
misalignment angles $\Delta\varepsilon$ and $\Delta\gamma$.
However the phase shift due  to the empty aluminum cell is two
orders of magnitude larger than the gas.  If the cell was present
in only one of the interferometer beams, a small unknown
misalignment would lead to a large systematic uncertainty in the
background phase shift. To minimize this systematic effect a
unique design was chosen for the gas sample cell, as discussed in
the next section.

\subsection{Gas cell design}
\label{sect:GasCellDesign}
In order to minimize the phase shift due to the cell, the cell
walls were designed to extend across both beam paths to produce
compensating phase shifts for paths $I$ and $II$. Mechanically
this was achieved by machining two gas cells from a single block
of aluminum.  The design of the cell is shown in
Fig.~\ref{fig:InterferometerNGasCell}.  When the cell is perfectly
aligned the beams strike both cell compartments perpendicular to
their surfaces. This ensures that the effective thicknesses along
both paths are nearly equal. The phase shift due to the cell
obtained from Eqs.~\ref{eq:phase} and \ref{eq:Deff2} is,
\begin{eqnarray}
    \label{eq:cellshift}
    \Delta\phi_{\mathit{cell}}(\varepsilon,\gamma) 
     & = & -\lambda N b \frac{(D_{II}-D_{I})}
           {\cos(\Delta\varepsilon)\cos(\Delta\gamma)}
\end{eqnarray}
where $D_{I}$ and $D_{II}$ are the cell wall thicknesses along paths
$I$ and $II$.  From Eq.~\ref{eq:cellshift} we see that the phase shift
along path $I$ is opposite that of path $II$ so that the total phase
shift is minimized when $D_{I}$ and $D_{II}$ are equal or nearly
equal, as is the case.  In this experiment the total phase shift of
the cell was first measured before filling with D$_{2}$ gas to be
$(\Delta\phi_{\mathit{cell}})_{D_{2}}~=~(2.4794\pm0.0021)$~rad and
$(\Delta\phi_{\mathit{cell}})_{H_{2}}~=~(1.3788\pm0.0021)$~rad.  The
two cell phase shift measurements differed by a slight amount due to
thin ($\approx10~\mu$m thick) film of thermal grease that was present
during the the D$_{2}$ gas measurement, but was removed (with acetone)
prior to the H$_{2}$ measurement.  These cell phase shift values are
considerably lower than 670~rad, which would have been seen with the
cell in only one beam.  The relative uncertainty introduced by a
17~mrad ($1^{\circ}$) misalignment of the cell is 0.001~\%.  This is
an order of magnitude below the uncertainty goal of this experiment. 
In practice, it is possible to align the cell in the beam such that
both $\Delta\varepsilon < 2$~mrad and $\Delta\gamma < 2$~mrad so that
the systematic alignment uncertainty is completely negligible.

\subsection{Alignment of the cell}
\label{sect:GasCellAlignment}
Although the experiment is insensitive in first order to both of the
missets $\Delta\varepsilon$ and $\Delta\gamma$, it is still necessary
to actually measure these values.  The fact that the cell was designed
to be extremely insensitive to changes in these angles introduces some
experimental difficulties in actually performing these measurements. 
Aligning the cell experimentally requires measurement of the cell
phase shift as a function of both $\Delta\varepsilon$ and
$\Delta\gamma$.  To solve this problem a kinematic mount was designed
to allow the cell to maintain the previous alignment
$\Delta\varepsilon$ and $\Delta\gamma$ relative to the beam after
being removed and replaced.  The alignment phase shift measurement was
performed using a quartz alignment phase plate that only crossed one
beam in order to produce a noticeable phase shift.  (See
Fig.~\ref{fig:InterferometerNGasCell}.)  This optically flat, 1.5~mm
thick quartz plate was mounted on the same flat surface as the cell. 
The procedure that was used to align the quartz is much the same as
the alignment procedure described in Ref.~\cite{Iof98} and is
described below.  However, here the sample is rotated by $\pi/2$
(90$^{\circ}$) relative to the non-dispersive phase shift position
discussed in \cite{Iof98}.

\begin{figure}[t]
    \begin{center}
    \includegraphics[width=2.764in]{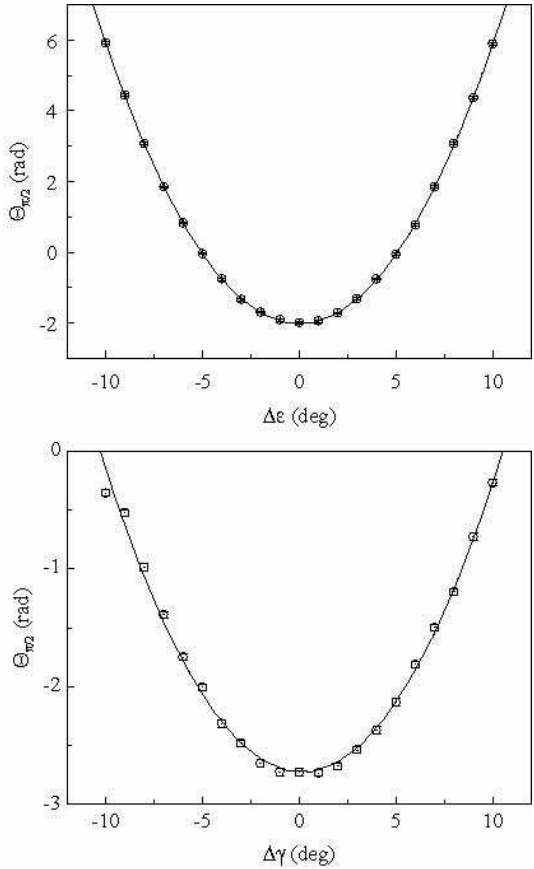}
    \end{center}
    \caption{Measurement of the difference phase $\Theta_{\pi/2}$
(Eq.~\ref{eq:phasediff}) to obtain the minimum of the rotation
alignment
    $\Delta\varepsilon$ and tilt alignment $\Delta\gamma$ as
    described in the text.}
    \label{fig:AlignParabola}
\end{figure}

This method of alignment required that the quartz alignment sample
be parallel translated from path $I$ to path $II$.  Upon
translation the horizontal misset angle changes sign
$(\Delta\varepsilon \rightarrow -\Delta\varepsilon$) yielding a
difference between the two phase shift measurements of
\begin{eqnarray}
    \label{eq:phasediff}
    \Theta_{\pi/2}(\varepsilon,\gamma) & = &
    \Delta\phi(\varepsilon,\gamma) - \Delta\phi(-\varepsilon,\gamma)\\
    & \approx & \frac{\lambda N b
D_{0}}{\cos(\Delta\gamma)\cos(\theta_{B})} \nonumber \\
    & & \times \left\{ 2 + (\Delta\varepsilon)^{2} \left[ 1 + 2
    \tan^{2}(\theta_{B})\right]\right\}. \nonumber
\end{eqnarray}
Note that when $\Delta\gamma$ is held fixed,
Eq.~\ref{eq:phasediff} is a quadratic function of
$\Delta\varepsilon$, which means that the center can be accurately
determined in order to minimize the misset angle
$\Delta\varepsilon$.  Similarly Eq.~\ref{eq:phasediff} is a
quadratic function of $\Delta\gamma$ when $\Delta\gamma<<1$ and
$\Delta\varepsilon$ is held fixed.  Independently varying the tilt
for a fixed value of $\Delta\varepsilon$ allows the experimental
determination of the optimum tilt position for $\Delta\gamma=0$.
The experimental measurements of these two alignment parabolas are
plotted in Fig~\ref{fig:AlignParabola}.  All phase shift data were
taken with $\Delta\varepsilon$ and $\Delta\gamma$ at their minimum
values.

\subsection{Data Collection}
\label{sect:DataCollection}
\begin{figure}[t]
    \begin{center}
    \includegraphics[width=2.867in]{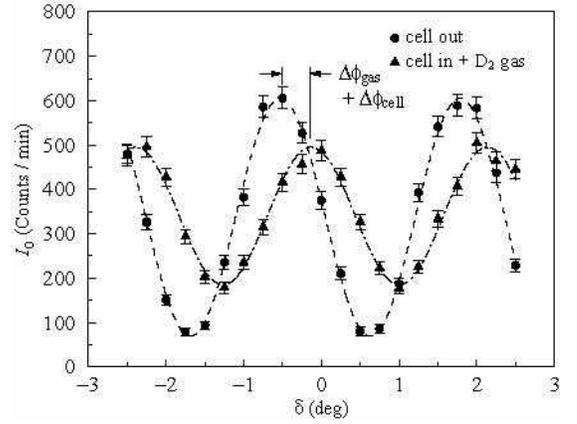}
    \end{center}
    \caption{A typical pair of interferograms denoting the change in
    intensity as the phase flag angle $\delta$ is varied (see
    Fig.~\ref{fig:Interferometer}).  Data are shown for both the cell
    filled with D$_{2}$ gas in the beam and when the cell is removed
from
    the beam path.}
    \label{fig:Interferogram}
\end{figure}

\begin{figure}[t]
    \begin{center}
    \includegraphics[width=2.793in]{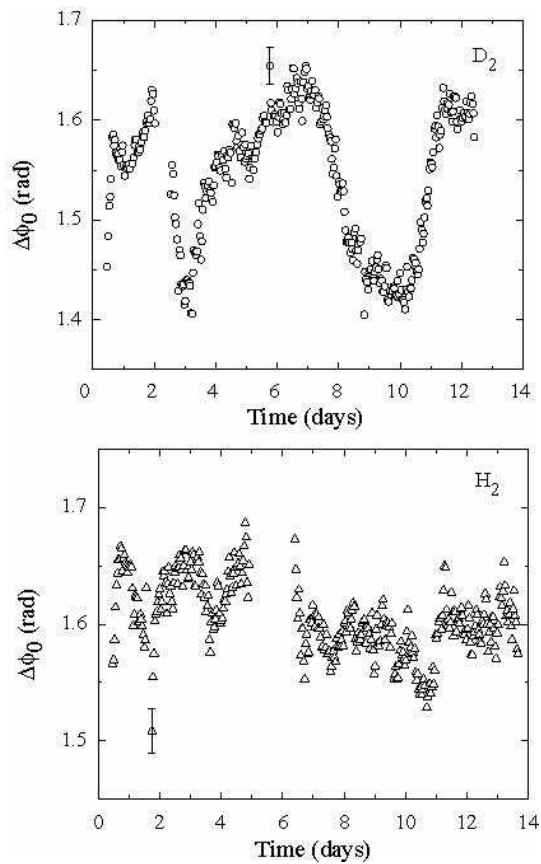}
    \end{center}
    \caption{Empty interferometer (cell removed) phase shift,
    $\Delta\phi_{0}$, plotted for each D$_{2}$ (upper) and H$_{2}$
    (lower) run.}
    \label{fig:PhaseOut}
\end{figure}

\begin{figure}[t]
    \begin{center}
    \includegraphics[width=2.793in]{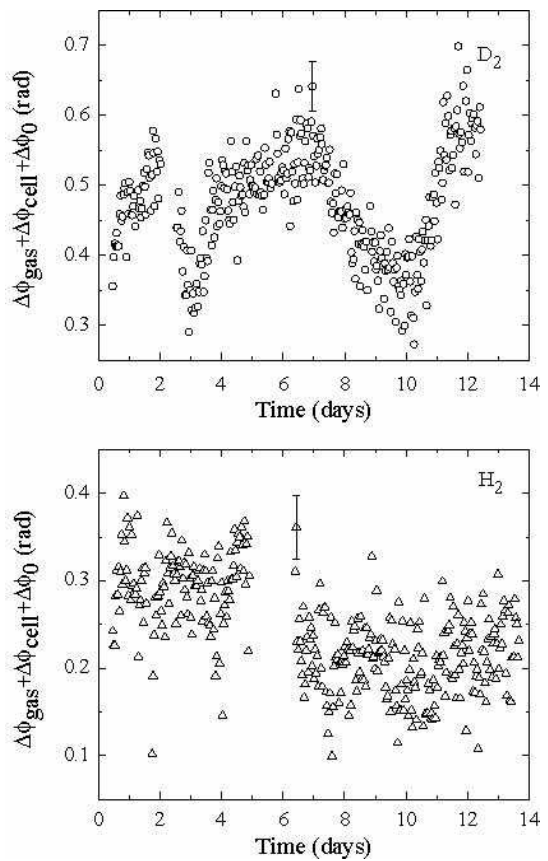}
    \end{center}
    \caption{Phase shift for the D$_{2}$ (upper) and H$_{2}$ (lower)
    filled gas cells as a function of measurement time.}
    \label{fig:PhaseIn}
\end{figure}

\begin{figure}[t]
    \begin{center}
    \includegraphics[width=3.12in]{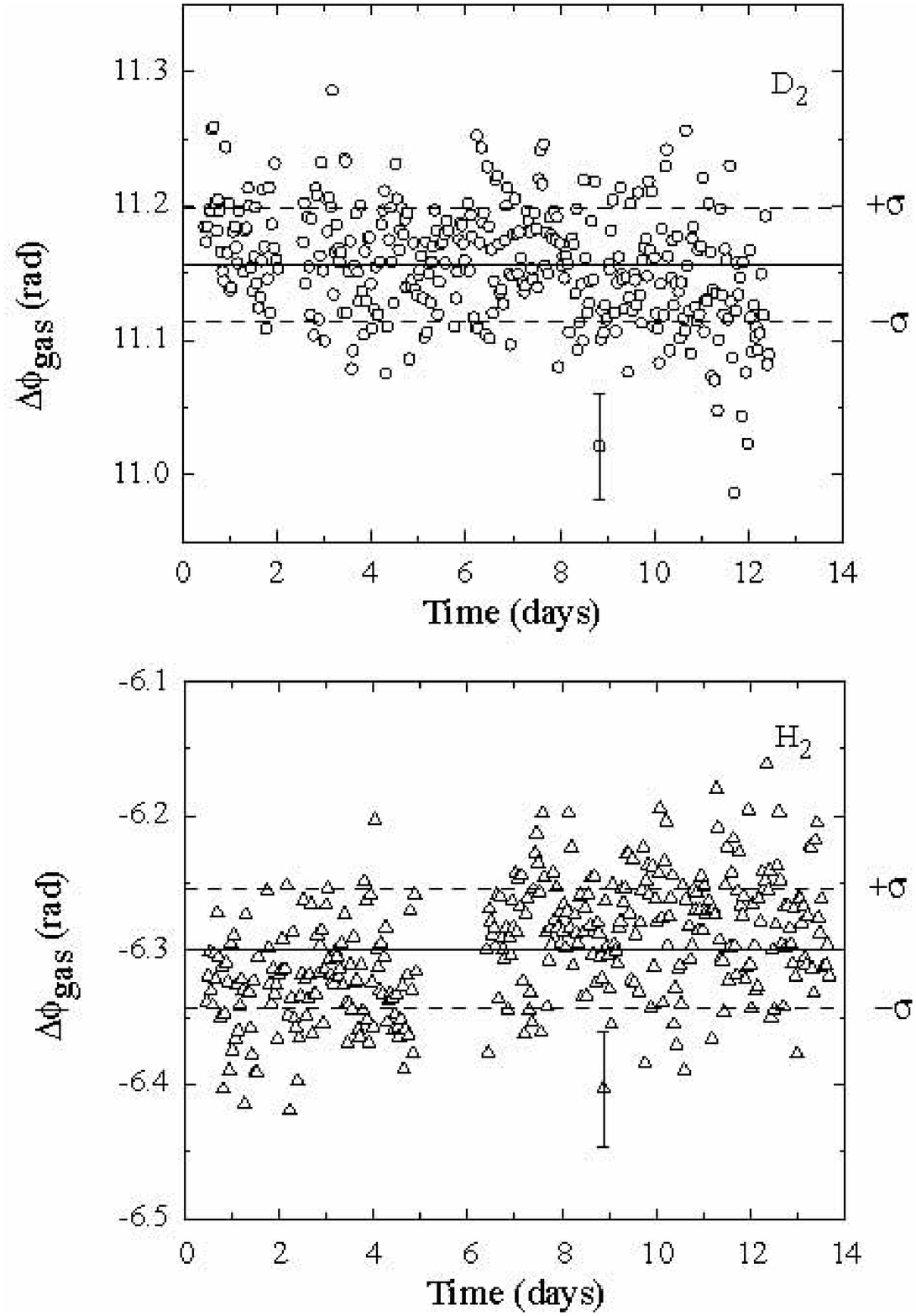}
    \end{center}
    \caption{Phase shift with the time-dependent initial phase shift
    subtracted off for (upper) D$_{2}$ and (lower) H$_{2}$ gas.  The
    mean values and their first standard deviation are shown in the
    figure.  One point in each figure is shown with the uncertainty
    estimated from Poisson counting statistics.}
    \label{fig:PhaseDif}
\end{figure}

Once the cell mount was aligned using the above technique, the
phase plate was replaced by the gas cell on its kinematic mount.
The gas chamber on path $I$ was then filled with sample gas, while
the compensation cell oriented symmetrically on path $II$ was
evacuated. Interferograms with the cell first in the ``cell in''
position and then translated to the ``cell out'' position (see
Fig.~\ref{fig:InterferometerNGasCell}) were collected in order to
determine the phase shift due solely to the gas and the small
difference in thickness between the two chamber walls of the cell
(discussed earlier in section~\ref{sect:GasCellDesign}). Each set
of interferograms (see Fig.~\ref{fig:Interferogram}) required
approximately 42~min to obtain a measurement of
$\Delta\phi_{\mathit{gas}}+\Delta\phi_{\mathit{cell}}$ with a
relative uncertainty of 0.3~\%. The two interferograms in
Fig.~\ref{fig:Interferogram} were fitted to Eqs.~\ref{eq:IO} and
\ref{eq:IH} to obtain the time dependent phase shifts,
$\Delta\phi_{0}$, shown in Fig.~\ref{fig:PhaseOut}. The total
phase shifts of the cell, the gas, and the time dependent offset
phase shifts,
$\Delta\phi_{\mathit{gas}}+\Delta\phi_{\mathit{cell}}+\Delta\phi_{0}$,
are plotted in Fig.~\ref{fig:PhaseIn}.  Since each measurement of
$\Delta\phi_{0}$ was performed within 42~min of the previous one,
the time dependence of the phase shift could be directly measured
and removed from the total phase shift.  The time dependence of
the empty interferometer phase shift, discussed later in
section~\ref{sect:Wavelength}, is believed to be due to slight
temperature fluctuations that cause small geometric shifts between
the two paths of the interferometer. The phase shift values with
the time dependence removed are plotted in
Fig.~\ref{fig:PhaseDif}. The total amount of data taken, 353 runs
for D$_{2}$ and 358 runs for H$_{2}$, was based on a statistical
uncertainty target of 0.02~\%.

\subsection{Measuring the atom density $N$}
\label{sect:AtomDensity}
To obtain a value for the coherent scattering length from the
phase shift data plotted in Fig.~\ref{fig:PhaseDif} using
Eq.~\ref{eq:phase}, several additional measurements must be
performed and a few time-dependent density changes must be
corrected for. First, it is necessary to determine the atom
density, $N$, of the gas for each phase shift measurement. To do
this we employ the following form of the virial equation for a
real diatomic gas given by,
\begin{eqnarray}
    \label{eq:gaslaw}
    N~=~\frac{2P}{k_{B}T(1 + B_{P}(T)P + C_{P}(T) P^2 + ...~)},
\end{eqnarray}
where $P$ is the pressure, $k_{B}$ is Boltzmann's constant (value
taken from \cite{Moh00}), $T$ is the temperature, $B_{P}$ is the
second pressure virial coefficient and $C_{P}$ is the third
pressure virial coefficient. (Note that the extra factor of two in
Eq.~\ref{eq:gaslaw} is included to account for the fact that each
molecule contains two atoms.) The virial coefficients for hydrogen
and deuterium have been measured \cite{Mic59,Dym80} with
sufficient accuracy to determine $N$ with a relative uncertainty
of 0.001~\% (for example: for deuterium
$B_{P}=5.616\times10^{-4}~$bar$^{-1}$,
$C_{P}=2.58\times10^{-7}$~bar$^{-2}$ and for hydrogen
$B_{P}=5.796\times10^{-4}$~bar$^{-1}$,
$C_{P}=2.36\times10^{-7}$~bar$^{-2}$ at T=297.15~K). The
correction to $N$ due to $B_{P}$ is about 0.7~\%, which is a
significant contribution. The correction due to $C_{P}$ is about
0.004~\% and is included in the calculation of $N$ for
completeness.

\begin{figure}[t]
    \begin{center}
    \includegraphics[width=2.915in]{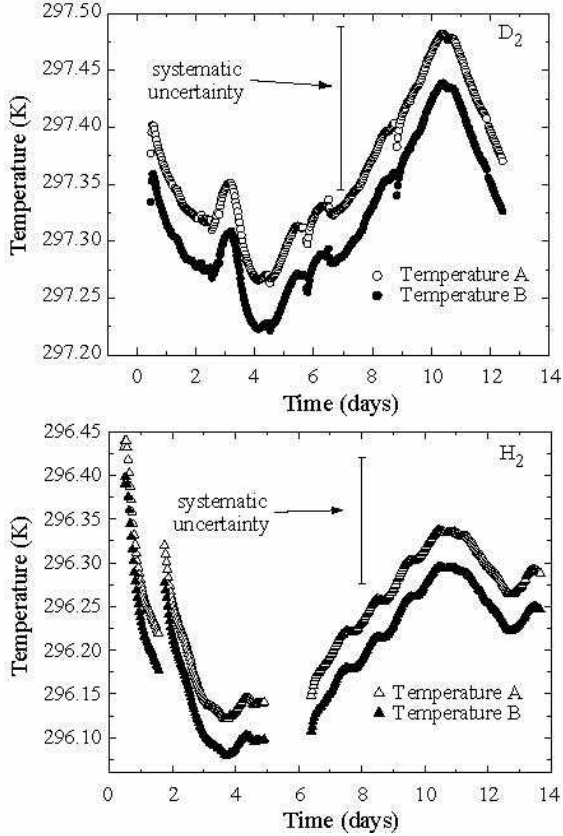}
    \end{center}
    \caption{The temperature of the cell measured with probes A and B
    (see text and Fig.~\ref{fig:InterferometerNGasCell}) for each data
    run of the D$_{2}$ (upper) and H$_{2}$ (lower) gas samples.  The
    temperatures of the two probes agree within the absolute accuracy
    of the measurement.  There is no evidence for a thermal gradient
    in the cell.}
    \label{fig:TemperatureData}
\end{figure}

To determine $N$ using Eq.~\ref{eq:gaslaw} both the absolute
temperature and the absolute pressure are needed.  The absolute
temperature of the gas was measured each time that the gas pressure
was measured.  To determine the absolute temperature, two calibrated
$100~\Omega$ platinum resistance thermometers were placed on the gas
cell at the positions labelled Temp A and Temp B in
Fig.~\ref{fig:InterferometerNGasCell}.  Although these thermometers
have a precision of $0.002~\%$, the electronic readout accuracy was
limited to $0.023~\%$ at 300 K. These thermometers were calibrated at
NIST by measuring the triple point temperature of pure H$_{2}$O.
Before absolute calibration, probe A reported the triple point
temperature as $(273.057\pm0.070)$~K and probe B reported
$(273.046\pm0.070)$~K, consistent within the absolute accuracy.  Since
the triple point of pure H$_{2}$O is the operational definition of
273.15~K absolute temperature, a shift of $(0.10\pm0.07)$~K was
applied to the temperature scales of both thermometers.  The small
differential nonlinearity of the platinum thermometers allows us to
apply the same shift at the 300~K temperature of the measurements.  A
plot of the temperature variation of the two probes using the original
probe calibration curves corrected for the absolute calibration using
the triple point is shown in Fig.~\ref{fig:TemperatureData}.  Note
that for both gases the temperature of the two cells was the same to
within the absolute accuracy of the thermometry indicating that there
is no thermal gradient across the cell.  Since the thermal
conductivity of the aluminum cell is high and there are no heat
sources or sinks other than the surrounding environment the
temperature of the gas can be inferred from the temperature of the
aluminum cell by assuming thermodynamic equilibrium.  The data show
there are slow temperature fluctuations: the amplitude of these
fluctuations is consistent with the rated temperature stability of the
interferometer enclosure.

The pressure was measured using a Mensor DPG II model 15000
\cite{disclaimer} digital pressure gauge.  This pressure gauge uses a
silicon pressure transducer (SPT), which modifies the output from a
precision micro-machined silicon wafer used as a pressure sensor.  It
was calibrated by the Mensor Corporation \cite{disclaimer} using a
dead weight test with NIST traceable standards.  The dead weight test
involves putting NIST traceable weights on a piston filled with dry
nitrogen.  The pressure exerted by the weight can be determined
absolutely by knowing the masses and the cross sectional area of the
cylindrical piston and compared with the output of the SPT.
Calibration measurements with the dead weight established that the SPT
readings were repeatable and linear functions of absolute pressure to
better than 0.01~\% \cite{Men00}.

\begin{figure}[t]
    \begin{center}
    \includegraphics[width=3.056in]{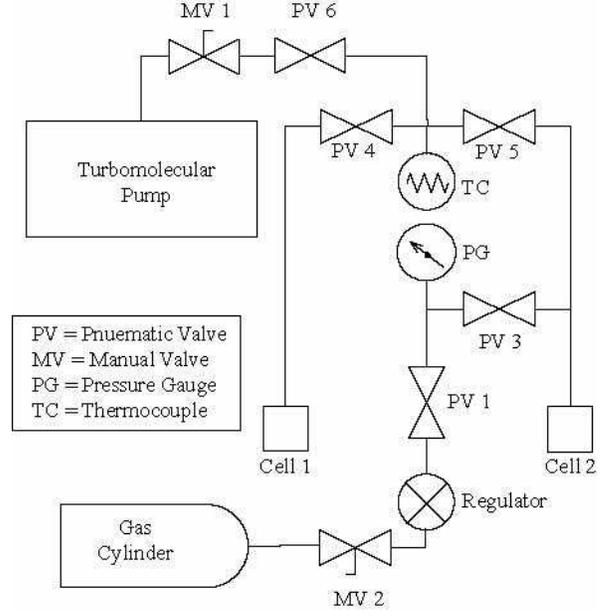}
    \end{center}
    \caption{Schematic of gas handling system used in experiments. 
    During data collection valves PV3, PV4, PV6, and MV1 were open.}
    \label{fig:gassys}
\end{figure}

This pressure gauge was directly coupled to the sample cell through a
high-pressure gas handling system schematically illustrated in
Fig.~\ref{fig:gassys}.  The top plate of the cell, which connects to
the body through an indium o-ring seal, was connected to the gas
handling system through two flexible 1~m long gas and vacuum lines
made from HiP~\cite{disclaimer} thin stainless steel tubes 1.59~mm OD,
and 0.76~mm ID. The thin flexible lines also allowed the cell to be
thermally insulated while allowing it to be translated in and out of
the interferometer without having to move the gas handling system. 
All connections in the gas handling system were made with all metal
seals to minimize contamination of the sample.

\begin{figure}[t]
    \begin{center}
    \includegraphics[width=2.845in]{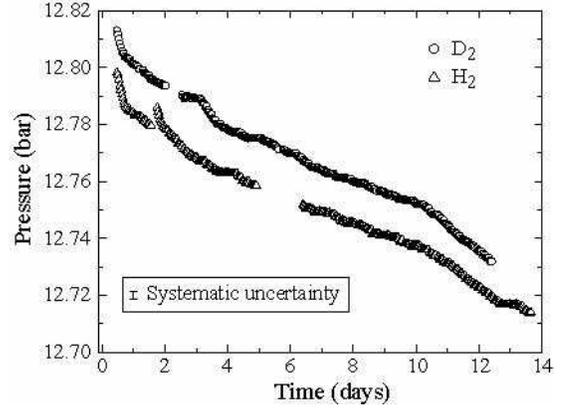}
    \end{center}
    \caption{Pressure of the D$_{2}$ and H$_{2}$ gas as a function of
    time.  The pressure decreases slowly due to an internal leak into
    the evacuated side of the cell. A small correction for this effect
    is incorporated into the data analysis.}
    \label{fig:PressureData}
\end{figure}

\begin{figure}[t]
    \includegraphics[width=2.853in]{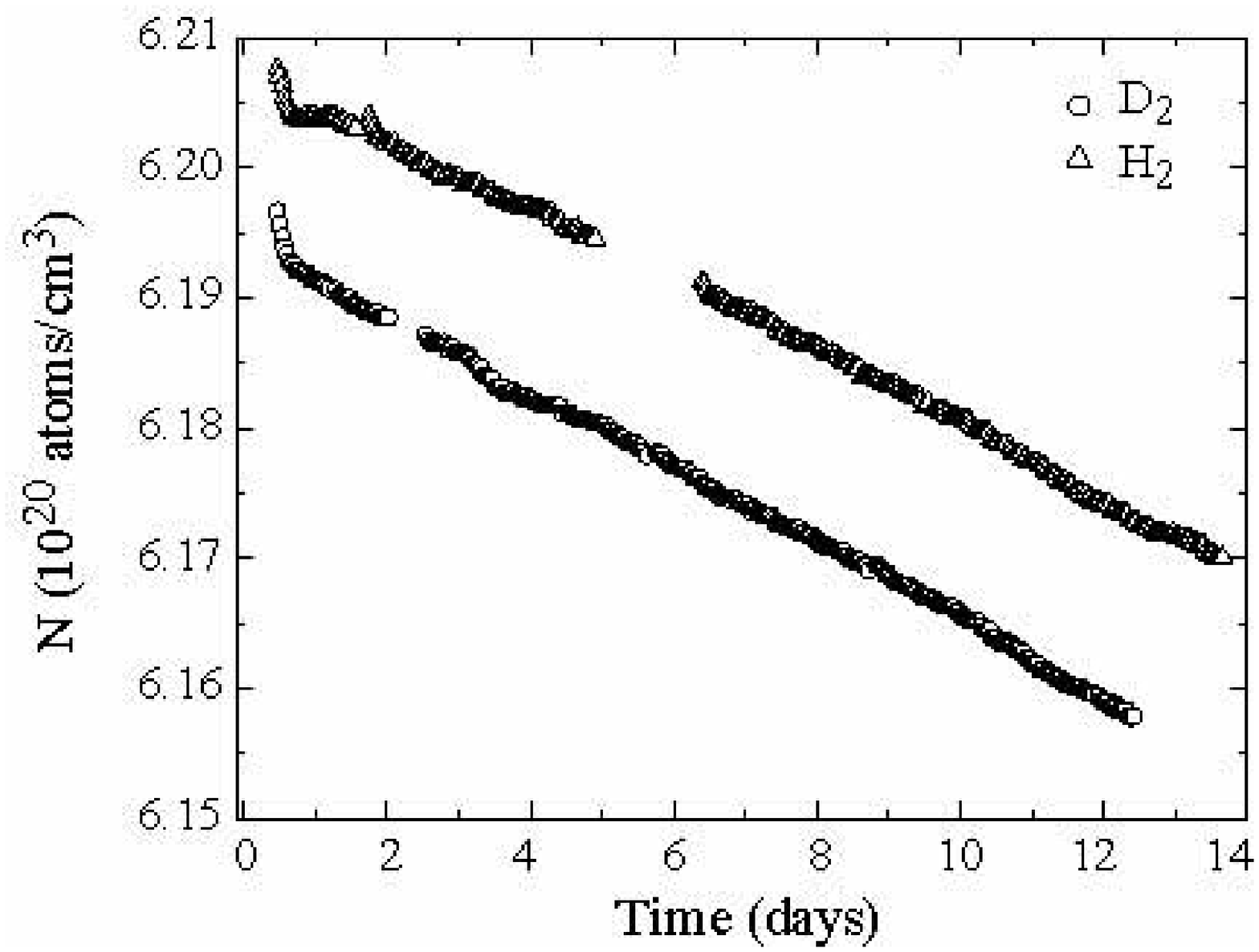}
    \caption{Atom density, $N$, of the D$_{2}$ gas and H$_{2}$ gas as
    a function of time.  The atom density is calculated from the virial
    equation using the run-to-run pressure and temperature values
    Figs.~\ref{fig:TemperatureData}~and~\ref{fig:PressureData}.}
    \label{fig:AtomDensity}
\end{figure}

The pressure shown in Fig.~\ref{fig:PressureData} decreases with time. 
This pressure change is not correlated with the temperature change of
the cell nor is it due to an external leak into to the system from
atmosphere.  It is also not due to diffusion of the gas into aluminum:
the known rate of diffusion of hydrogen into aluminum is much too slow
\cite{Ichi92}.  This leak appeared only when one cell was evacuated
and the other cell was pressurized.  The conclusion drawn from this
information is that the cell leaked into the evacuated cell from the
filled cell through gas valve PV5 in Fig.~\ref{fig:gassys}.  Although
not an ideal condition, the presence of such an internal leak does not
compromise the purity of the gas.  Also, the evacuated side was
continuously pumped during the experiment so that there would not be
enough gas in the evacuated section to cause a systematic error in the
phase shift measurement.  The leak did change the atom density, $N$,
at a rate of $0.002~\%~$h$^{-1}$.  However, this slow change was
measured and can be corrected for easily in the data analysis.  With
data for both the temperature and pressure, the time-dependent atom
density $N$ is calculated from Eq.~\ref{eq:gaslaw}, with the results
shown in Fig.~\ref{fig:AtomDensity}.

\subsection{Gas Purity}
\label{sect:GasPurity}
The D$_{2}$ gas used in these measurements was ``Research Grade''
quality purchased from Air Products \cite{disclaimer}.  The quoted
purity from the manufacturer was 99.99~\%, but HD and H$_{2}$
impurities were not included in this analysis.  The H$_{2}$ gas was
``Ultra High Purity'' purchased from Matheson Tri-Gas
\cite{disclaimer} with quoted purity of 99.999~\%.

\begin{figure}[t]
   \includegraphics[width=3.146in]{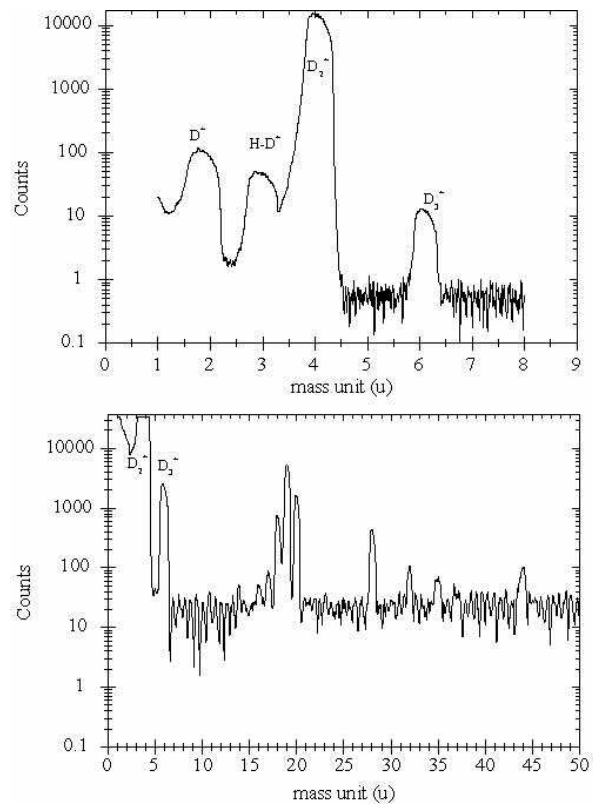}
    \caption{Low mass spectrum (upper) for D$_{2}$ gas showing the
    relative peak heights of D$^{+}$, HD$^{+}$, D$_{2}^{+}$ and
    D$_{3}^{+}$.  The full mass spectrum (lower) shows the D$_{3}^{+}$
    peak height relative to all the higher mass contaminants.  The
    ionization efficiencies of masses 18 thru 20 are about three times
    that of D$_{2}$.  Mass 19 (fluorine) is a background due to
    outgassing of teflon in the mass spectrometer.}
    \label{fig:MassSpect}
\end{figure}

Measurement of the actual impurity species and concentrations in
the gas samples used is important to correct the measured value of
$b$ at the 0.01~\% level. To do this both mass spectroscopy and
Raman spectroscopy were employed.  The mass spectra allowed one to
see the contaminating components in the samples as shown for
example for D$_{2}$ gas in Fig.~\ref{fig:MassSpect}.  From the
data, the main contaminants found in the D$_{2}$ gas were HD at
0.3~\% and D$_{2}$O at a 0.02(1)~\% level. However, the relative
ionization efficiencies of HD and D$_{2}$ are not known for the
mass spectrometer used in this measurement.  Therefore, to measure
the HD concentration to better than 5~\% accuracy, Raman
spectroscopy was used \cite{Comp93}.

\begin{figure}[t]
    \includegraphics[width=2.93in]{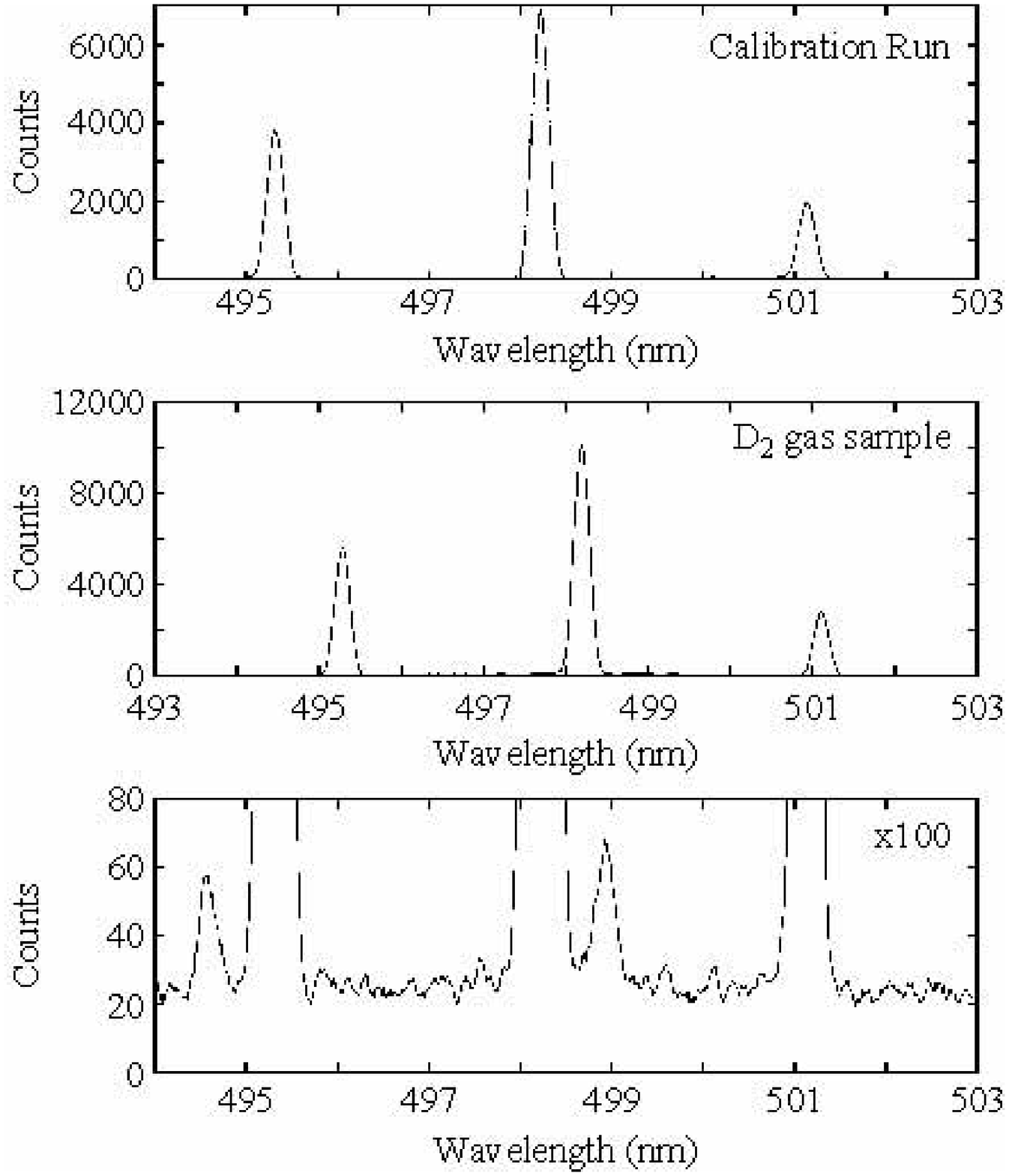}
    \caption{Raman spectroscopy of D$_{2}$ gas.  The upper plot is the
    calibration with a known HD ratio of 2~\%.  The middle and lower 
    plots are the
    rotational spectrum of the D$_{2}$ gas sample used in this
    experiment.  This determined the calibration of HD to D$_{2}$ peak
    in the mass spectrum data of contaminants.}
    \label{fig:RamanSpect}
\end{figure}

Raman spectroscopy measures the amount of light scattering from the
rotational levels of HD and D$_{2}$, which are shifted due to the
isotopic mass difference.  Measurements were made using the apparatus
described in Ref.~\cite{Liu02}.  The spectrometer was first calibrated
with a known 2~\% HD sample (see Fig.~\ref{fig:RamanSpect}a).  Next,
the D$_{2}$ gas was introduced into the apparatus and rotational
spectra were taken (Fig.~\ref{fig:RamanSpect}b,c).  From this data the
mole fraction of HD was determined to be $x_{HD}=(0.00301\pm0.00013)$. 
The accuracy (4~\%) of this measurement was limited by the sample
pressure and in principle could be lowered to 1~\% with minor changes.

With the Raman data for HD it is then possible to calibrate the
mass spectrum data in order to determine the amount of other
contaminants relative to D$_{2}$. The mole fractions of D, H, and
O were inferred to be $x_{D}=(0.99840\pm0.00017)$,
$x_{H}=(0.001500\pm0.000065)$, and $x_{O}=(0.000050\pm0.000016)$.
The final expression for $b_{D}$, corrected for impurities, was
obtained using the following relation:
\begin{equation}
    \label{eq:scatlengthcorr}
    b_{D}  =   \frac{b_{gas} - b_{H} x_{H} - b_{O} x_{O}}{x_{D}}
\end{equation}
where $b_{H} = (-3.7410\pm0.0020)$~fm and $b_{O} =
(5.805\pm0.004)$~fm \cite{Koe91} were used.


\subsection{Sample thickness}
\label{sect:Thickness}
\begin{figure}[t]
   \includegraphics[width=3.026in]{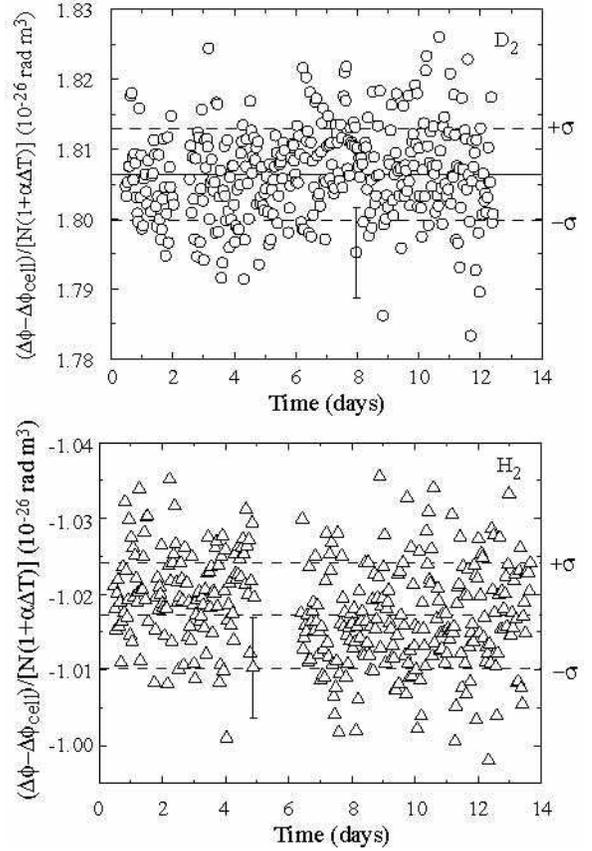}
    \caption{Point-to-point gas phase shift divided by the
    point-to-point atom density, $N$, and the point-to-point
    correction for the length of the cell due to thermal expansion and
    the temperature difference between the neutron measurement and the
    length measurement.  The mean values and their first standard
    deviation are shown in the figure.  One point in each figure is
    shown with the uncertainty estimated from Poisson counting
    statistics.  The final standard uncertainty ($\sigma /
    \sqrt{N-1}$) of the mean must be added in quadrature to the
    systematic uncertainties due to the pressure and temperature
    measurement.  }
    \label{fig:ReducedPhase}
\end{figure}

Measurement of the sample thickness was performed using the NIST
Precision Engineering Division Coordinate Measuring Machine (CMM)
\cite{Stoup}.  This device is a temperature controlled coordinate
measuring apparatus capable of measuring macroscopic distances with an
uncertainty of 250~nm within a 98~\% confidence interval.  This
measurement showed that the thickness of the gas filled chamber of the
cell was uniform at the 0.01~\% level and that the cell was
$(1.0016\pm0.0001)$~cm thick at an absolute temperature of
$(20.00\pm0.05)~^{\circ}$C. Variation of the temperature resulted in a
systematic change in the thickness at the 0.01~\% level during the
scattering length measurement.  Using the coefficient of linear
expansion for aluminum $\alpha=2.5\times10^{-5}~^{\circ}\mbox{C}^{-1}$
\cite{CRC}, this variation was taken into account in order to obtain
the ratio $\Delta\phi / [N(1+\alpha\Delta T)]$, where $\Delta T$ is
the difference between the temperature measured in the experiment and
the temperature at which the cell dimensions were measured by the NIST
CMM. This ratio is plotted in Fig.~\ref{fig:ReducedPhase} from which a
mean value and a standard error of the mean was calculated.

\begin{figure*}
    \includegraphics[width=4.71in]{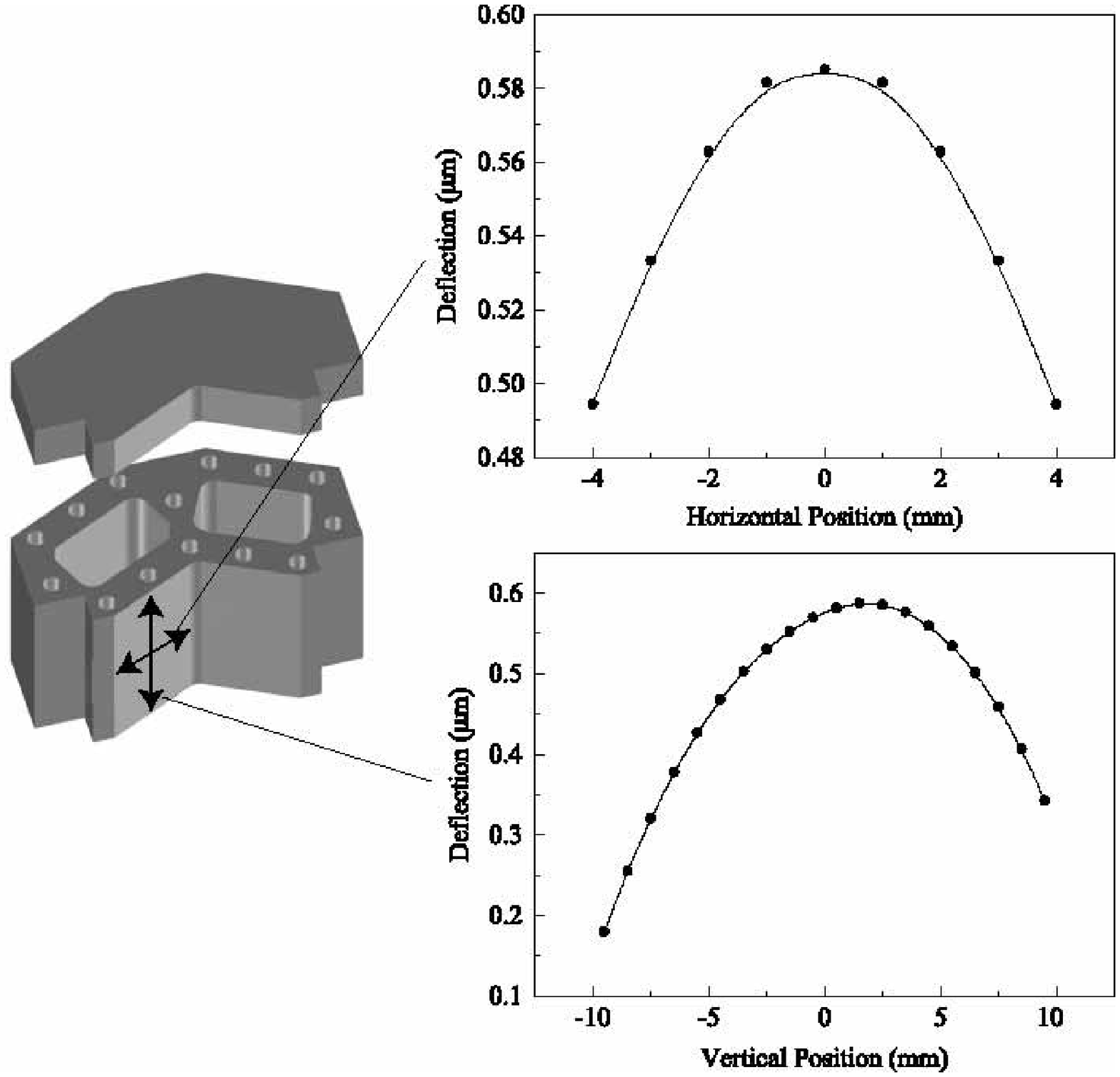}
    \caption{The three-dimensional perspective of the gas cell
    modelled using finite element analysis is shown.  Although the
    entire surface was mapped in the actual calculation, only the
    central and horizontal axes (where the maximum deflection occurs)
    are shown for a pressure of 13~bar.}
    \label{fig:FiniteElement}
\end{figure*}

Dimensional changes in the cell also occurred when the cell was
pressurized to $\approx12.8$~bar, which caused the walls to flex
slightly outward.  An upper limit for this effect was determined
by both measurement using a Mitutoyo indicator \cite{disclaimer}
with 1~$\mu$m resolution and calculation using finite element
analysis (see Fig.~\ref{fig:FiniteElement}). Both calculation and
measurement confirmed that this flex resulted in a change in
thickness at the center of the cell of less than 1~$\mu$m, which
amounts to a systematic effect on the thickness of less than
0.01~\%.

The fact that the time dependence of the measured phase shifts shown
in Fig.~\ref{fig:ReducedPhase} has disappeared after all the known
time-dependent effects on the gas density are accounted for is a
nontrivial result.  The odds that other time-dependent effects not
accounted for are present and conspire to cancel are very low, and
there are no plausible physical mechanisms for other time-dependent
effects on the phase shift beyond the ones mentioned so far
(time-dependent wavelength shifts were eliminated as a possibility as
described in the next section).  Since the coherent scattering length
being measured is a time-independent constant, establishing the
time-independence of the phase shift is a necessary condition for a
valid measurement.

In addition, the statistical spread of the measured phase shifts is
consistent with that expected based on Poisson statistics and the
statistical accuracy of the individual phase shift measurements based
on neutron counting statistics.  This confirms that there are no
unknown sources of noise present in the measurement and places further
indirect constraints on possible interfering influences on the data.

\subsection{Measuring $\lambda$}
\label{sect:Wavelength}
The wavelength was measured using an analyzer crystal in the H-beam of
the interferometer shown in Fig.~\ref{fig:Interferometer}.  Rotating
the analyzer crystal through both the symmetric and the anti symmetric
Bragg reflections allows the absolute Bragg angle, $\theta_{B}$, to be
determined, and thereby the wavelength, using Bragg's law,
$\lambda~=~2~d~\sin(\theta_{B})$, where $d$ is the lattice plane
spacing of the crystal.  Analyzer crystals of both pressed Si and
pyrolytic graphite (PG) were used to determine $\theta_{B}$.  Silicon
is ideal since the lattice constant is known with an uncertainty much
lower than our requirement of 0.01~\%\cite{Bas94}.  The Si crystal was
used at first along with the PG crystal to allow the PG lattice
constant to be measured.  The PG crystal was mounted kinematically to
the interferometer setup to ensure that the neutron beam was always
sampled by the same part of the PG crystal.  Therefore each
measurement of the wavelength is correlated with the original
measurement performed using silicon.

\begin{figure}[t]
    \includegraphics[width=2.973in]{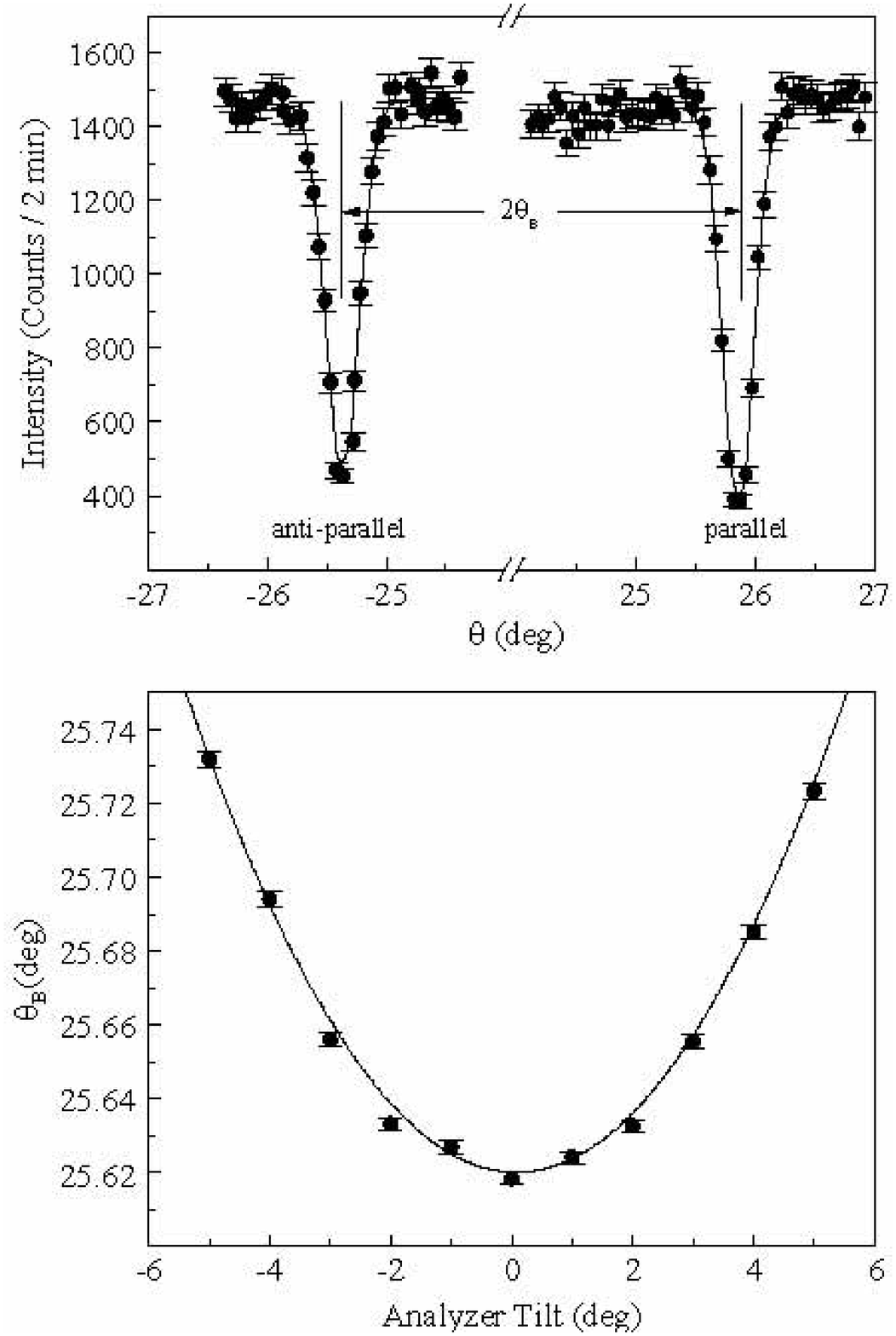}
    \caption{A typical set of rocking curves (intensity vs.  angle)
    for the analyzer crystal at a fixed tilt angle (above).  The
    $2\theta_{B}$ angle is calculated from the difference between the
    center of the parallel Bragg curve and the anti-parallel curve. 
    Below, the $2\theta_{B}$ angles are plotted as a function of the
    tilt angle of the analyzer .  The minimum of the parabola
    corresponds to the correct analyzer tilt position, and therefore
    the correct Bragg angle $\theta_{B}$.}
    \label{fig:WavelengthAlign}
\end{figure}

The crystal was rotated about the vertical axis to locate the two
angles where the Bragg condition is satisfied.  This condition is
manifested as a dip in the transmitted beam measured in the $^{3}$He
detector.  These rocking curves (intensity versus angle) are fitted to
Gaussians to determine the centroids of the dips
(Fig.~\ref{fig:WavelengthAlign}).  The wavelength, obtained from the
$2\theta_{B}$ angle was determined from the angular difference between
the centroids.  By performing measurements of the $2\theta_{B}$ angle
with both Si and PG with the same neutron wavelength the lattice
constant of PG can be calibrated relative to that of Si.

For the Si measurements it was necessary to include an additional
linear term in the fit to account for a changing background due to a
parasitic Bragg reflection of reciprocal lattice points that lie near
the Ewald sphere of reflection.  This effect does not appear in the
data for PG since the lattice of PG is randomly oriented about the
hexagonal c-axis.  (This random orientation effectively makes the
reciprocal lattice of the PG crystal one-dimensional.)

\begin{figure}[t]
    \includegraphics[width=2.955in]{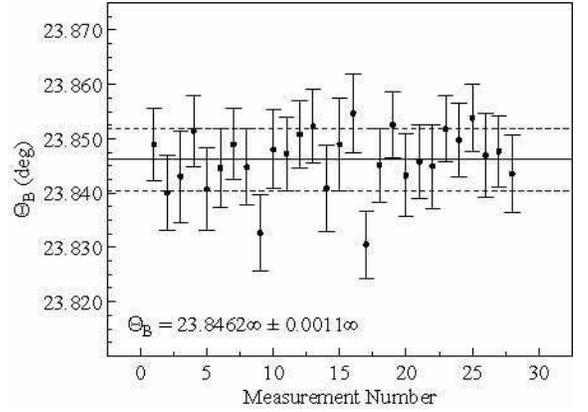}
    \caption{Set of Bragg angle values for a set of measurements of
    the type shown in Fig.~\ref{fig:WavelengthAlign}.  Note that the
    data is shown for the calibrated PG (002) analyzer crystal at the
    optimum tilt position.  The value of the PG lattice parameter,
    $c$, calibrated against the Si-lattice parameter ($a =
    0.543101993$~nm) is $c = 0.670982$~nm.}
    \label{fig:WavelengthData}
\end{figure}

The measurement of the $2\theta_{B}$ angle is also sensitive to the
relative tilt between the interferometer lattice planes and the
analyzer lattice planes.  In order to align the analyzer crystal,
measurements of $2\theta_{B}$ were performed for various tilt angles
of the analyzer lattice planes.  The functional form of $2\theta_{B}$
plotted versus the tilt angle is a parabola (see
Fig.~\ref{fig:WavelengthAlign}) with a minimum corresponding to the
condition that the lattice planes of the analyzer crystal are
perpendicular to the scattering plane of the interferometer.  This
method allows $\theta_{B}$ to be determined with an uncertainty of
0.001~\%.  The resulting value for $\lambda$, determined from the data
shown in Fig.~\ref{fig:WavelengthData}, was found to be
$(0.271266\pm0.000012)$~nm for the D$_{2}$ gas experiment and
$(0.2713050\pm0.0000085)$~nm for the H$_{2}$ gas experiment.  These
two numbers differ slightly at the 0.01~\% level due to minor changes
in the mechanical configuration of the apparatus between the
experiments.  These changes were due to the need to remove the cell
between the measurements.

Not only is it necessary to know the mean value of the wavelength
of the beam passing through the interferometer, but it is also
necessary to ensure that this value does not change during the
course of the experiment.  Unlike the temperature and pressure
measurement the wavelength was measured only twice during the
experiment for practical reasons. To prove that the time-dependent
phase drift described previously was not due to time-dependent
wavelength changes a thick aluminum sample was placed in the beam
which amplified the phase shift in the interferometer by a factor
of 100, thereby amplifying any fluctuations in the phase due to
possible wavelength changes by the same factor.  The result of
this test showed that as long as no mechanical changes in the
experimental configuration have been made, the long-term stability
of the wavelength is better than 0.001~\%.  This means that,
within limits of the statistical uncertainty, the phase drift is
due to temperature fluctuations and not due to a drift of the
wavelength.

\section{Results for $b_{D}$ and $b_{H}$}
\label{sect:ExperimentalResults}

All of the measurements required to calculate the coherent
scattering lengths in H$_{2}$ and D$_{2}$ gas are now determined.
It remains to relate the coherent scattering lengths of the
molecules, $b_{H}$, and $b_{D}$, to the coherent scattering
lengths of the nuclei, $b_{\mathit{np}}$, and $b_{\mathit{nd}}$.

The general expression for the low energy neutron scattering
length of an atom away from nuclear resonances for unpolarized
atoms and neutrons is
\begin{equation}
    \label{eq:alllengths}
    b=b_{nuc}+Z(b_{ne}+b_{s})[1-f(q)]+b_{s}+b_{pol}
\end{equation}
where $b_{nuc}$ is the scattering amplitude due to the
neutron-nucleus strong force, $b_{ne}$ is the neutron-electron
scattering amplitude due to the internal charge distribution of
the neutron, $b_{s}$ is the Mott-Schwinger scattering due to the
interaction of the magnetic moment of the neutron with the ${\bf
v} \times {\bf E}$ magnetic  field seen in the neutron rest frame
from electric fields, $b_{pol}$  is the scattering amplitude due
to the electric polarizability of the neutron in the intense
electric field of the nucleus, and $f(q)$ is the charge form
factor (the Fourier transform of the electric charge distribution
of the atom). We first note that the electromagnetic contribution
to the scattering lengths of both hydrogen and deuterium from
$b_{ne}$ and $b_{s}$ are exactly zero for forward scattering due
to the neutrality of the atoms, which forces the charge form
factor $f(q) \to 1$ as $q \to 0$ \cite{Searsbook}. In addition,
the contribution from the electric polarizability of the neutron
for both H and D is less than $-0.000017$~fm\cite{Sch91} and is
therefore negligible. Thus this measurement is sensitive only to
that part of the scattering length due to the strong interaction,
which is precisely what can be  calculated using theoretical
models of the NN interaction. Although  there are in principle
additional contributions due to local field effects that arise
from multiple scattering in the medium and its modification of the
amplitude of the incident waves on the scattering centers, these
corrections are completely negligible for gases and only reach the
$10^{-4}$ level in solids and liquids\cite{Searsbook}. Therefore
both our measurement, and the most accurate of the past
measurements, measure the same physical quantity  and therefore
can be directly compared to each other as well as to NN theory.

A summary of the parameters discussed in the previous section are
given in Table~\ref{tab:parameters}. Using these values as input
to Eq.~\ref{eq:phasediff} we obtain the average molecular bound
coherent scattering length per atom for deuterium of \bD\ and for
hydrogen of \bH.

\begin{figure}[t]
    \includegraphics[width=3.144in]{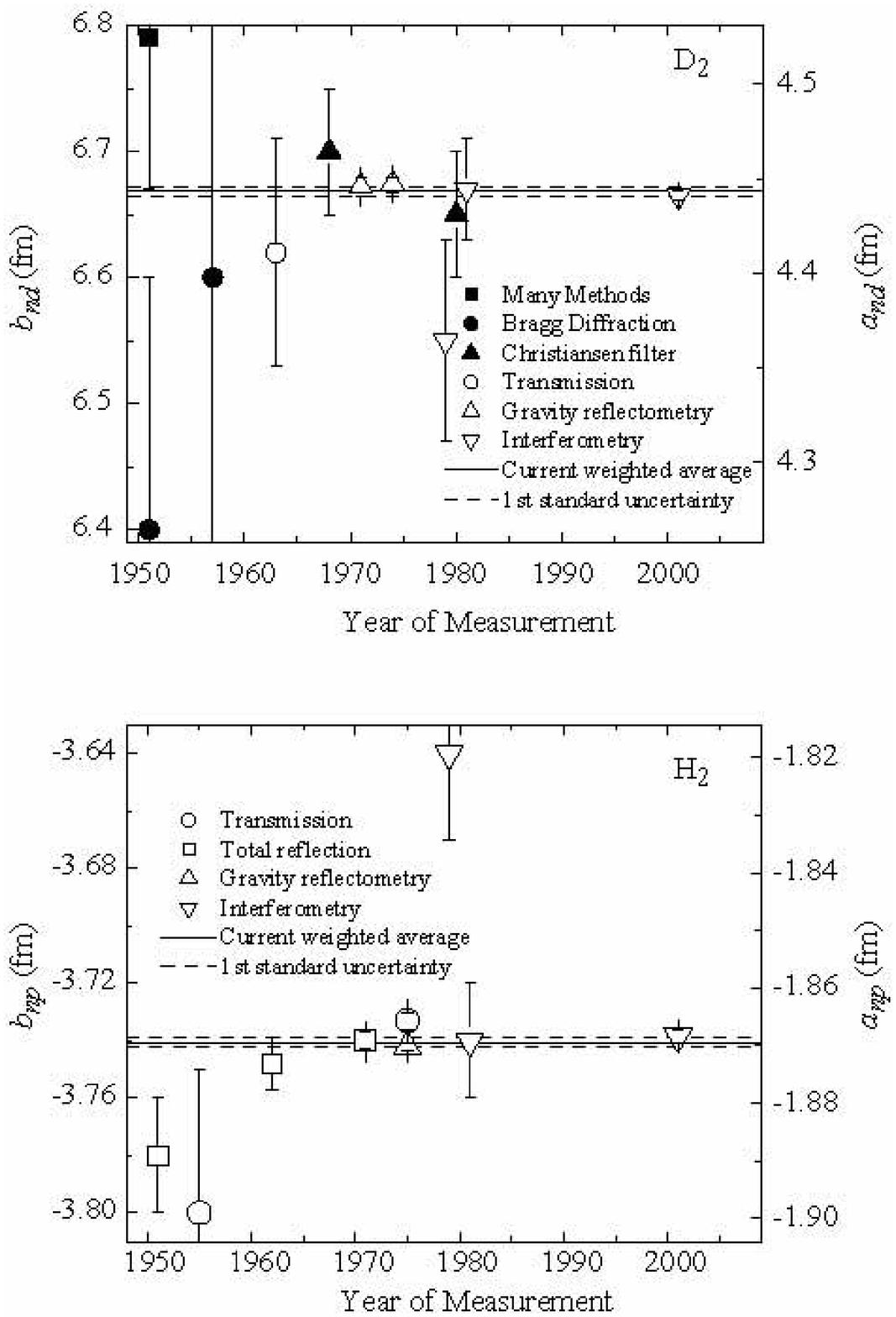}
    \caption{Bound coherent neutron scattering lengths for n-d and n-p
    along with reported uncertainties.  Our result for D$_{2}$ is
    consistent with the current world average.  Our result for
    H$_{2}$, uncorrected for the multiple scattering and second-order
    perturbation theory corrections discussed in the text, is
    $2.3\sigma$ away from the current world average.  Including these
    theoretical corrections brings our value into closer agreement as
    shown.}
    \label{fig:ScatteringLengthHistory}
\end{figure}

To compare this result with other measurements we must calculate
an average value from the previous world values of the
$b_{\mathit{np}}$ and $b_{\mathit{nd}}$ bound nuclear coherent
scattering lengths.  The details of this calculation are left to
Appendix~\ref{sect:WorldAverageCalculation} and the result of this
evaluation shown in Fig.~\ref{fig:ScatteringLengthHistory} is
\bnpprevworld\ and \bndprevworld. Our result differs from the
average of previous measurements by $2.3\sigma$ for H and
$1.9\sigma$ for D.

\begin{table*}
    \caption{Parameters and relative
    uncertainties required to determine the scattering length.  The
coherent
    scattering length is calculated using the expression in
Eq.~\ref{eq:phase},
    the parameters in this table, and the correction for D$_{2}$ gas
composition
    in Eq.~\ref{eq:scatlengthcorr}.}
    \label{tab:parameters}
    \begin{tabular}{|l|l|l|l|l|}
        \hline
        Parameter           & Value(D$_{2}$) & Relative
$\sigma$(D$_{2}$)
        & Value(H$_{2}$) & Relative $\sigma$(H$_{2}$)  \\
        \hline
        $\Delta\phi$ & 13.64~rad    & $1.9\times 10^{-4}$
        & -4.9192~rad    & $3.5\times 10^{-4}$  \\
        \hline
        $\Delta\phi_{\mathit{cell}}$ & 2.4794~rad    &
$8.4\times10^{-4}$
        & 1.3788~rad    & $1.5\times 10^{-3}$  \\
        \hline
        $\lambda$           & 0.271266~nm    & $4.4\times 10^{-5}$ &
        0.271305~nm    & $3.1\times 10^{-5}$  \\
        \hline
        Temperature         & see data       & $2.3\times 10^{-4}$ &
see
        data       & $2.3\times 10^{-4}$  \\
        \hline
        Pressure         & see data       & $1.0\times 10^{-4}$
        & see data       & $1.0\times 10^{-4}$  \\
        \hline
        $D_{\mathit{cell}}$          & 1.0016~cm      &
$1.0\times10^{-4}$
        & 1.0016~cm      & $1\times 10^{-4}$  \\
        \hline
        $\Delta\phi_{gas}/[N(1+\alpha\Delta T)]$
        & $1.80650\times 10^{-26}$ rad m$^3$ & $2.6\times 10^{-4}$
        & $-1.01787\times 10^{-26}$ rad m$^3$ & $4.7\times 10^{-4}$\\
        \hline
        x D$_{2}$           & 0.9968         & $1.6\times 10^{-4}$ &
        0.0000         & 0.0000  \\
        \hline
        x HD                & 0.0030         & 0.043
        & 0.0000         & 0.0000  \\
        \hline
        x D$_{2}$O          & 0.00020        & 0.5
        & 0.0000         & 0.0000  \\
        \hline
        x H$_{2}$           & 0.0000         & NA
        & 1.0000         & $1\times 10^{-4}$  \\
        \hline
        x H$_{2}$O          & 0.0000         & 0.0000
        & 0.0000         & 0.0000  \\
        \hline
        $b_{atom}$ & \bdmolourvale    & \bdmolourunc & \bpmolourvalue  & \bpmolourunc  \\
        \hline        \NowakSymbol & 0.0000~fm      & 0 & 0.0070~fm     & 0  \\
        \hline
        $b_{nuclear}$ & \bdcorourval      & \bdcorourunc & \bpcorourval  & \bpcorourunc   \\
        \hline
        $b_{prev~world~avg}$ & \bdprevworldval      & \bdprevworldunc & \bpprevworldval  & \bpprevworldunc   \\
        \hline
        $b_{curr~world~avg}$ & \bdcurworldval      & \bdcurworldunc & \bpcurworldval  & \bpcurworldunc   \\
        \hline
    \end{tabular}
\end{table*}

This comparison is not yet complete since there remains one class
of effects, which although not generally taken into account in
past measurements, can in principle be large enough to introduce a
nontrivial correction in the usual relation between the scattering
length of the molecule and the scattering lengths of the
constituent atoms. In 1982, Nowak \cite{Now82b} revisited the
approximations  inherent in the use of the Fermi pseudopotential
to describe the refractive index, $n$, for neutrons, which had
first been investigated by Lippman \cite{Lip50b} and was the first
application of the Lippmann-Schwinger formulation of scattering
theory\cite{Lip50a}. Nowak came to the conclusion that the static
local field corrections \cite{Searsbook} were of the order of
$10^{-4}$, in agreement with Lax's earlier estimates \cite{Lax52}.
However, he found that dynamic effects due to virtual excitations
to the low lying states, within second order perturbation theory,
make a relative contribution of order $10^{-3}$ to $(1 - n)$.  The
effect is most pronounced for light target molecules such as
H$_{2}$ and presumably for this reason have rarely been taken into
account in n-p coherent scattering length measurements, most of
which involve measurements in which the hydrogen is bound in
hydrocarbons. (These corrections are also mentioned in the n-p
parahydrogen cross section measurements of Callerame \cite{Cal75},
which is one of the measurements used to determine the n-p
scattering length.) The lowest lying excitations of the molecule
are rotational states, $E_{rot} = (\hbar^{2}/2I)(j)(j+1)$, where
$I$ is the molecule's moment of inertia. For H$_{2}$,
$(\hbar^{2}/2I) = 7.56$~meV; the vibrational energy levels are
separated by 546~meV. Nowak carried out a detailed numerical
calculation of this dynamic correction, finding it to be a
relative correction of $1.1~\times~10^{-3}$. A later estimate by
Summerfield \cite{Sum84} agrees with this conclusion. Nowak also
calculated the correction to $(1 - n)$ due to multiple scattering
of the neutron within the H$_{2}$ molecule. This leads to a
relative correction of about $0.8~\times~10^{-3}$. Both of these
corrections are spin-dependent so that the para- and ortho-states
of the H$_{2}$ molecule must be considered separately. The
absolute correction to the scattering length calculated from the
sum of these two relative corrections is \NowakCorrectionH, which
is consistent with the difference between our neutron
interferometry measurement of $b_{\mathit{np}}$ and the previous
world average calculated using $b_{H}=b_{\mathit{nd}}+\Delta
b_{Nowak}$.  This previous world average includes results from
reflectometry measurements of hydrocarbon liquids. It is presumed
that the measurements of $b_{\mathit{np}}$ for hydrogen bound in a
hydrocarbon sample will not need to be corrected for multiple
scattering or virtual excitations due to the inverse molecule mass
dependence of the effect.

This measurement of $b_{\mathit{np}}$ by neutron interferometry on
H$_{2}$ gas may therefore be regarded as the first observation of
the corrections to the index of refraction due to virtual
excitations and multiple internal scattering within a molecule as
predicted by Nowak 20 years ago. This conclusion must be regarded
as tentative at this point, since the calculation of Nowak was
done in the long wavelength limit $kR_{0} \rightarrow 0$ ($R_{0}$
= bond length of H$_{2}$ = 0.74611 nm, for D$_{2}$ $R_{0}$ =
0.74164 nm \cite{CRC}). We are currently performing a calculation
for the conditions of our experiment, which correspond to
$kR_{0}$=1.73.

If we apply the existing Nowak correction to our H$_{2}$ data, our
result for the coherent n-p scattering length is \bnp. Taking this
value we calculate the new world average value to be \bnpcurworld,
(our result lies within $1\sigma$ of the average).  The Nowak
correction is expected to be much larger for H$_{2}$ than for
D$_{2}$ due to the fact that there is a sign difference between
the singlet and triplet scattering lengths of hydrogen, which
results in an amplification of the correction to the scattering
amplitude given by Eq.~\ref{eq:hydrocorrection}.  The consequence
of this sign difference is most prominently seen in the the
incoherent scattering cross section of hydrogen, which as a
consequence is disproportionately larger than the incoherent
scattering cross section of deuterium. Therefore, we do not make a
similar Nowak correction to the D$_{2}$ data.  With this
uncorrected value of $b_{\mathit{nd}}$ we arrive at a new weighted
average of \bndcurworld\ (our result is within $0.9\sigma$ of the
average value). Our results are consistent with past measurements
conducted decades earlier by completely different methods.

\section{Comparison to Theory and Implications for Three-Body Forces}
\label{sect:CompWithTheory}
We propose to show in this section that the physics impact of
these coherent scattering length measurements in the n-d system
are greater than has been recognized in the theoretical few body
community. The reason is simple: rather than compare theoretical
calculations to the coherent scattering lengths, which are
measured to high precision, theorists in the field have instead
compared their results to the scattering lengths in each of the
two channels individually despite the fact that these are known
with a precision that is lower by an order of magnitude.

To illustrate the impact of comparing theoretical calculation to
the precision data on the coherent scattering length, which has
actually existed now for decades, we show several modern
calculations of the scattering lengths  in the p-d and n-d
systems, some of which include a three-nucleon force. The
dependence of the theoretically calculated $^2$S$_{1/2}$
scattering length on the inclusion of a 3N force is clearly seen
in Table~\ref{TheoryHistory}, which gives the n-d and p-d doublet
and quartet scattering lengths (in fm) calculated with a number of
different potential models. The results in boldface come from
models in which the 3N force has been adjusted to replicate the
trinucleon binding energy.

\begin{table}
\caption{Theoretical calculations of the p-d and n-d $s$-wave
scattering lengths. In the last column we have calculated the
coherent scattering lengths for the n-d system by forming the
appropriate linear combination. We observe that none of the
theories, with the possible exception of the AV14 potential with a
3N force, are in agreement with the precisely-known world average
coherent n-d scattering length of $(6.669\pm0.003)$~fm.}
\label{TheoryHistory}
\begin{tabular}{llrrrrr}
{\bf Ref.} & \multicolumn{1}{c}{{\bf Potential}} &
\multicolumn{1}{c}{$^{2}a_{\mathit pd}$} &
\multicolumn{1}{c}{$^{4}a_{\mathit pd}$} &
\multicolumn{1}{c}{$^{2}a_{\mathit nd}$} &
\multicolumn{1}{c}{$^{4}a_{\mathit nd}$} &
\multicolumn{1}{c}{$b_{\mathit nd}$}\\
\mbox{} & \multicolumn{1}{c}{{\bf model}} &
\multicolumn{1}{c}{[fm]} & \multicolumn{1}{c}{[fm]} &
\multicolumn{1}{c}{[fm]} & \multicolumn{1}{c}{[fm]} &
\multicolumn{1}{c}{[fm]} \\
\hline \hline
\cite{Fri83} & Yukawa          & -3.6  & 13.6   & -2.1  & 6.39  & 5.34 \\
\cline{2-6} 
\mbox{}      & Exponential     & -7.6  & 13.4   & -6.4  & 6.44  & 3.24 \\
\cline{2-6} 
\mbox{}      & {\bf MT I--III} & 0.15  & 13.8   & 0.70  & 6.44  & 6.79 \\
\hline
\cite{Fri84} & RSC-5           & 2.23  & ---    & 1.76  & ---   & ---  \\
\cline{2-6} 
\mbox{}      & AV14            & 1.42  & 13.57  & 1.35  & 6.38  & 7.06 \\
\cline{2-6} 
\mbox{}      & SSCC            & 1.35  & 13.67  & 1.32  & 6.41  & 7.07 \\
\cline{2-6}  
\mbox{}      & {\bf RSC-5}\footnote{for this potential, the $^3$S-wave NN 
potential has been adjusted to replicate the trinucleon binding energies} 
                               & 0.06  & ---    & 0.60  & ---   & ---  \\
\hline 
\cite{Che91} & RSC             & 1.569 & 13.55  & 1.52  & 6.302 & \\
\cline{2-6} 
\mbox{}      & RSC+TM3NF       &-0.509 & 13.568 & 0.393 & 6.308 & 6.505 \\
\cline{2-6} 
\mbox{}      & AV14            & 0.967 & 13.764 & 1.200 & 6.372 & 6.972 \\
\cline{2-6} 
\mbox{}      & AV14+BR3NF      &-1.133 & 13.764 & 0.001 & 6.378 & 6.378 \\
\cline{2-6} 
\mbox{}      & {\bf RSC+TM3NF} & $\approx$ 0.0 
                                       & 13.52  &  0.66 & 6.30  & 6.63 \\
\cline{2-6} 
\mbox{}      & {\bf AV14+BR3NF}& $\approx$ 0.0  
                                       & 13.76  &  0.57 & 6.38  & 6.67 \\
\hline 
\cite{Ber86} & Yamaguchi       & 0.257 & 13.68  & 0.656 & 6.27  & 6.60 \\
\hline 
\cite{Che89} & {\bf MTI--III}  & 0.17  & 13.8   & 0.71  & 6.43  & 6.79 \\
\hline 
\cite{Kie94} & {\bf MTI--III}  & 0.003 & 13.96  & 0.702 & 6.442 & 6.793 \\
\cline{2-6} 
\mbox{}      & AV14            & 0.954 & 13.779 & 1.196 & 6.380 & 6.798 \\
\hline 
\cite{Kie97b}& AV14            & ---   & ---    & 1.189 & 6.379 & 6.974 \\
\cline{2-6} 
\mbox{}      & {\bf AV14+TM3NF}& ---   & ---    & 0.5857& 6.371 & 6.664\\
\hline 
\cite{Kie95a}& {\bf AV14+BR3NF}& -0.178& ---    & 0.575 & ---   & ---  \\
\cline{2-6} 
\mbox{}      & {\bf AV18+UR3BF}& -0.022& ---    & 0.626 & ---   & ---  \\
\hline
\cite{Kie97a}& \mbox{}         & 0.024 & 13.8   & ---   & ---   & ---  \\
\hline \hline
\end{tabular}
\end{table}

\begin{figure}[t]
    \includegraphics[width=2.45in]{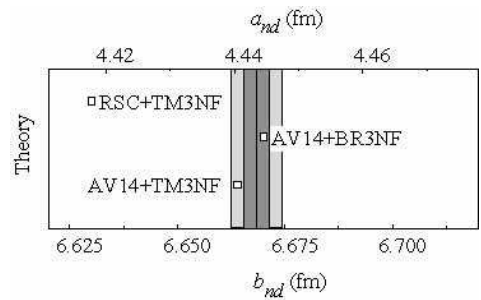}
    \caption{Theoretical calculations of the coherent scattering
    length compared with the experimental value measured here.  The
    central dark band is the 1$\sigma$ confidence band and the lighter
    band is the 2$\sigma$ confidence band.  Only one of the theories
    fall within the 1$\sigma$ band and only 2 fall within 2$\sigma$.}
    \label{fig:TheoryCoherent}
\end{figure}

In addition, the values of the coherent scattering lengths from
Table~\ref{TheoryHistory} are plotted in Fig.~\ref{fig:TheoryCoherent}
along with the world average value for $b_{\mathit{nd}}$ with
$1\sigma$ and $2\sigma$ confidence bands.  We note that, as expected,
none of the theories which do not incorporate a 3N force of some sort
come close to matching the n-d coherent scattering length.  Of the
potential models that include a 3N force of some type, only the AV14
potential with the Brazil 3N force\cite{Che91} and the AV14 potential
with the Tucson-Melbourne 3N force\cite{Kie97b} are in agreement with
the data.  The AV18 potential with the Urbana 3N force, the AV14
potential with the Brazil 3N force, the Malfliet-Tjon-III potential,
the Reid soft-core+Tuscon-Melbourne 3N force, the Reid soft-core
adjusted to fit the triton binding energy, and none of the other
potential models which include the Brazil 3N force\cite{Che91,Kie94}
are in agreement with the data on the n-d coherent scattering length
or, in cases where there is no separate calculation of the quartet n-d
scattering length, in agreement with the doublet n-d scattering length
derived below.  This indicates that the precision with which the
coherent n-d scattering length is known sets a tight constraint on NN
potential models and, in our view, ought to be considered as
automatically as is the value of the triton binding energy when such
potentials are compared to the low energy observables.  With the new
result of the analysis of n-d scattering in chiral effective field
theory\cite{Epe02} including chiral 3N forces which shows that only
two low energy parameters are required at NNLO order (surprisingly
none are required at NLO\cite{Kol94}), the question of which
observables to use to fix these low energy constants becomes timely. 
We believe that the best two low energy observables in the n-d system
to determine these low energy constants are {\bf not} the triton
binding energy and the doublet n-d scattering length but rather the
triton binding energy and the coherent n-d scattering length.

We can also use the coherent scattering length data in combination
with theoretical calculations of the quartet scattering length to
infer the doublet scattering length with significantly higher
precision than the currently quoted value of \DoubletExpPrev.  As is
well-known, the quartet scattering length is mainly sensitive to the
well-known long-range components of the NN interaction due to pion
exchange, which are fixed by measurements of the s-wave component of
the deuteron wave function.  So the results of NN model calculations
should give the same answer for this channel to high accuracy
independent of the details of the short-range components of the NN
interaction where 3-body forces start to manifest themselves.  Then
this procedure can be used to more tightly constrain the short-range
interactions, including 3-body forces, which must be introduced to
agree with the triton binding energy and to calculate the doublet
scattering length.

Table~\ref{TheoryRecentQuartet} shows calculations of the quartet n-d
scattering length using a new class of potentials which provide good
fits to the NN database\cite{Fri00}.  We observe that all of these
results fit within a range \QuartetTheory.  If we accept this average
and range as a fair representation of the precision with which modern
NN potentials can calculate $^{4}a_{\mathit{nd}}$, then we can combine
this result with our measurement of the coherent scattering length to
obtain the following value for the doublet scattering length:
\DoubletExpTheoryNew.  Compared with the direct measurement
\DoubletExpPrev, we see that this approach can improve on the
precision of our knowledge of the doublet n-d scattering length by a
factor of 4.
\begin{table}
\caption{Theoretical calculations of the quartet n-d $s$-wave
scattering length. Table taken from Friar \etal\cite{Fri00}}
\label{TheoryRecentQuartet}
\begin{tabular}{|l|l|}
        \hline
        Potential  & $^{4}a_{\mathit{nd}}$  \\
        \hline
        \bf {N93}\cite{Sto94} & 6.346 \\
        \hline
        \bf {NII}\cite{Fri93} & 6.343 \\
        \hline
        \bf {RSC93}\cite{Sto94} & 6.353 \\
        \hline
        \bf {CDB}\cite{Mac96} & 6.350 \\
        \hline
        \bf {AV18}\cite{Wir95} & 6.339 \\
        \hline
        \bf {Chiral PT}\cite{Bed98}& 6.33 \\
        \hline
\end{tabular}
\end{table}

At this point it would also be interesting to compare these new
results with the scattering lengths in the p-d system.
Unfortunately, there are at present {\em no} currently accepted
values for the low energy p-d scattering lengths. The analysis of
Black \etal, did reproduce some of the theoretically predicted
energy dependence of the p-d $^2$S$_{1/2}$ effective range
function--a singularity near threshold--but the singularity was
located at higher than expected energies and did not match the
theoretical values within quoted uncertainties\cite{Bla99}.  None
of the other experimentally determined values for this parameter
were even close to the calculated values, and the effective range
functions did not display the expected pole.  For the p-d case,
the conflict between theory and experiment persists whether or not
3N forces sufficient to correctly bind $^3$He are added to the
potential.

\section{Conclusions and Further Work} 
\label{sect:Conclusion}
We have performed high precision measurements of the coherent neutron
scattering lengths of gas phase molecular hydrogen and deuterium using
neutron interferometry.  We find \bH\ and \bD. Our result for H
differs from the world average of previous measurements,
\bnpprevworld, by \OurExpDifference, which is accounted for by the
Nowak correction.  Our result for the n-d coherent scattering length
is in agreement with the world average of previous measurements,
\bndprevworld.  We feel that the precision of these results has yet to
be properly appreciated.  We note that calculations of the doublet and
quartet scattering lengths of the best potential models show that
almost all known calculations are in disagreement with the
precisely-measured linear combination corresponding to the coherent
scattering length.  Combining the world data on $b_{D}$ with the
modern high-precision theoretical calculations of the quartet n-d
scattering lengths recently summarized by Friar \etal, we deduce a
more precise value for the doublet scattering length of
\DoubletExpTheoryNew.  This value is a factor of 4 more precise than
the previously accepted value of \DoubletExpPrev\ and is in agreement
with the Argonne AV18 potential with a 3N force.

We hope that this work will contribute to the extensive theoretical
and experimental efforts now underway to understand the nuclear 3-body
force.  When possible, we urge theorists who calculate both scattering
lengths in the n-d system to compare to the precisely known coherent
scattering length in addition to the lower-precision values of the
separate scattering lengths.  In trying to use the n-d system to
constrain possible forms of the 3-body force this procedure should be
more sensitive.  In particular, we hope that the n-d coherent
scattering length and the triton binding energy will be used to
constrain the two NNLO low energy constants that model 3N forces in
chiral effective field theory.

It would of course be very useful to have another precision
measurement in the n-d system which is sensitive to a different linear
combination of scattering lengths so that both scattering lengths
could be extracted from experiment alone with high precision. 
Performing a interferometry measurement with a polarized D$_{2}$
target of sufficient density to operate stably in a neutron
interferometer seems impractical at the moment due to the cryogenic
and high magnetic field requirements needed to make polarized D$_{2}$. 
Although polarized D$_{2}$ gas targets using spin-exchange optical
pumping are under development, the densities reached so far are still
too low for a precise neutron interferometer experiment\cite{Rholt}. 
One possibility is to perform a measurement of the pseudomagnetic
precession of polarized neutrons in a polarized D$_{2}$ target.  Such
a measurement determines a phase shift proportional to the difference
in the scattering lengths.  Recently such a high-precision measurement
was performed with polarized $^{3}$He at the ILL\cite{Zim02}.  It
would be very interesting if it were possible to adapt the elegant
neutron spin precession measurement technique used by Zimmer \etal\ in
this measurement to the case of polarized D or H.

Another natural question to ask is how the current measurements can be
improved if more neutron flux with a neutron interferometer were to
become available.  The absolute measurement of the gas density can be
improved by at least an order of magnitude before one needs to worry
about the accuracy of the virial coefficients.  A skew symmetric
interferometer with 10 cm beam paths could be constructed that would
allow the sample cell path to be increased by nearly an order of
magnitude reducing the relative dimensional uncertainty by that amount
as well.  Designing the interferometer to work at a longer wavelength
of 0.38 nm would also increase the sensitivity by a factor of 1.4. 
Measurement of the HD contamination of the gas in the case of D$_{2}$
could be improved from 4~\% relative uncertainty to 1~\%.  With all of
these improvements taken into account the uncertainty can in principle
be reduced by nearly an order of magnitude.

Finally it is worth noting that the last decade has witnessed exciting
advances in the accuracy of calculations in few-body nuclei. 
Therefore it is possible now to envision the accurate calculation of
low energy neutron scattering lengths for systems with $A>3$.  The
next most interesting systems to measure are clearly $^{3}$H,
$^{3}$He, and $^{4}$He, and as room temperature gases all the
techniques described in this paper are applicable to them.  In a
future paper we will present results of a precision measurement of the
n-$^{3}$He coherent scattering length, which in combination with the
recent measurement of the scattering length difference in the same
nucleus will be used to determine separately the scattering lengths in
both channels.  This should improve the accuracy of our knowledge of
low energy scattering observables in the 4-body system by an order of
magnitude.  It would also be possible to improve the current
measurements of the coherent neutron scattering length of tritium. 
Again the motivation would need to come from theoretical developments
in the 4-body problem.

Any NN potentials that correctly describe the two and three-body
systems, even if a three-body force needs to be added, must predict
correctly the n-$^{3}$He scattering length with no adjustable
parameters.  A deviation would indicate either the presence of
four-body forces (but these are believed to be even smaller than
three-body forces on theoretical grounds) or a distortion of the NN
interaction in the nuclear medium.  Recent theoretical calculations
exist for the n-$^{3}$He system\cite{Kie98}.

\section{Acknowledgements}
\label{sect:Acknowledgements}
We would like to thank the NIST Center for Neutron Research for
support and use of the neutron facilities during this experiment, Los
Alamos National Laboratory for the use of the Raman Spectroscopy
facilities, Philip Lawson for translating a paper for us from German,
and Dr.  William Dorko of the NIST Analytical Chemistry Division for
assistance with the mass spectroscopy measurement of the gas sample
composition.

This work was supported by the U.S. Department of Commerce, the NIST
Ionizing Radiation Division, the Department of Energy, the National
Science Foundation Grant No.  PHY-9603559 at the University of
Missouri, and National Science Foundation Grant No.  PHY-9602872 at
Indiana University.

\appendix
\section*{Appendix}
\subsection{Corrections to the Impulse
Approximation used in Neutron Optics and Consequences for High
Precision Scattering Length Determinations}
\label{sect:NowakCorrection}
We have chosen to describe in some detail the assumptions of the
usual theory of neutron optics in section~\ref{sect:NeutOptTheory}
because we need to consider an effect which violates some of the
approximations which underly this formalism. The approximation
fails in two places:  in Eqs. \ref{eq:tapprox} and
\ref{eq:FermiApprox}, due to effects from atomic binding and
multiple scattering respectively.

To understand the effects of atomic binding, it is worth recalling
the approximations which underly equation \ref{eq:FermiApprox}.
Since the neutron-nucleus interaction is much stronger and much
shorter range than the binding forces of the atoms in matter, it
is reasonable to neglect the effects of chemical binding during
the neutron-nucleus collision. In addition, the short-range of the
interaction means that the timescale of the collision is much
shorter than the timescales associated with the motion of the atom
in the potential well. For both of these reasons, the $t$ operator
is usually approximated by the $t$ operator for a free atom. This
is known as the impulse approximation in scattering theory.

In this approximation, the neutron optical properties of a medium
depend only on the coherent scattering length of the atoms and not
at all on the details of the binding of the atoms. This
independence from the medium has been exploited for decades in
precision measurements of neutron coherent scattering lengths,
which have typically been determined by measurements in mixtures
and compounds chosen for their experimental convenience. As a
matter of principle, however, at some point this approximation
will no longer be valid and binding effects will need to be taken
into account. When this regime is reached, it will be necessary to
know enough details about the atomic binding to calculate their
effects on the optical potential. It is our assertion that this
regime has now been reached in our experiment.

The calculation of corrections to the impulse approximation for
neutron scattering lengths was plagued for decades by the singular
nature of the delta function potential for the $t$
matrix\cite{Lip50b,Lip50b,Bre47a,Bre47b,Sum64}.  However in a series
of papers\cite{Now82b,Die81,Now82a} manifestly divergence-free
expressions for the first-order corrections to the impulse
approximation were finally obtained which are consistent with the
optical theorem and reduce in appropriate limits to previous results
in the limit of static scatterers\cite{Sea82}.  As an example, the
case of a nucleus of mass $M$ in an atom bound in a harmonic
oscillator potential, the first-order correction to the impulse
approximation was evaluated analytically\cite{Die81} with the result
\begin{equation}
    \label{eq:b_oscillator}
    f=-b\left[1-\frac{3bA}{2\alpha\sqrt{\pi}} \int_{0}^{\infty} ds 
    \frac{1-(1+s)e^{-s}}{(As+1-e^{-s})^{5/2}}\right],
\end{equation}
where $f$ is the scattering amplitude, $A=M/m$ is the ratio of the
nuclear mass to the neutron mass, $b$ is the bound scattering length,
and $\alpha$ is the average amplitude of the oscillator in the ground
state
 \begin{equation}
    \label{eq:amplitude}
    \alpha^{2}= \frac{\hbar }{ 2M\omega}
\end{equation}
with $\omega$ the oscillator frequency. The scale of the
correction is set by the ratio of the scattering length (a few fm)
to the vibration amplitude of the atom (0.1~nm or so), which is
small and typically neglected in neutron optics theory. However,
since the size of this correction is a decreasing function of $A$,
it is largest for H and D. In the particular case of a bound
proton the scattering amplitude becomes
\begin{equation}
    \label{eq:hydrocorrection}
    f= \frac{-1 }{ 2}(3b_{3}+b_{1})\left[1-0.578
    {\frac{(3b_{3}^{2}+b_{1}^{2}) }{ \alpha(3b_{3}+b_{1})}}\right]
\end{equation}
where $b_{3}$ and $b_{1}$ are the triplet and singlet n-p
scattering lengths and the numerical factor of $0.578$ comes from
evaluating the integral. The factor multiplying the expression is
the usual coherent scattering length for hydrogen. For the
particular case of hydrogen, where $b_{3}=5.42$~fm and
$b_{1}=-23.75$~fm, the size of the correction term is anomalously
large from the small denominator, $1.0\times10^{-3}$ for a typical
atomic oscillation amplitude of $0.05$~nm.

However there is another effect of the same order of magnitude
that must be taken into account associated with intraparticle
multiple scattering. In Eq.~\ref{eq:tapprox} the $t$ matrix of the
bound system of $N$ scatterers is expressed as the sum of the
(impulse approximation) one-body $t$ matrix. But this is known in
exact treatments of scattering theory to be an
approximation\cite{Gol64}. The next order of approximation for the
$t$ matrix of the system is
\begin{equation}
    \label{eq:multiplescat}
    t= \sum_{l}t_{l} + \sum_{l,l^{'}, l\ne l^{'}} t_{l}Gt_{l^{'}}+\ldots
\end{equation}
where $G$ is the Green's function. Therefore the $t$ operator for
the system to second order is not the sum of the $t$ operators for
the individual particles due to interparticle multiple scattering.
So to calculate the full $t$ operator of the system, which is what
is required to obtain the optical potential in
Eq.~\ref{eq:coherent} one must take multiple scattering also into
account. Then finally the optical potential, which is now a
function of the neutron momentum, must be solved from
Eq.~\ref{eq:tmatrix} and inserted into
Eq.~\ref{eq:refractionindex}.

Nowak\cite{Now82b} performed this calculation to obtain the
modified expression for the index of refraction
\begin{equation}
    \label{eq:newindex}
    n^{2}=1-\frac{v(k)}{E}
\end{equation}
and two correction terms at second order: one from the binding
potential
\begin{eqnarray}
    \label{eq:potentialcorrection}
    v_{b}^{(2)}=& A& \int_{- \infty}^{+ \infty}dt  
    \int d^{3}{\bf q} e^{\frac{i\hbar t }{ 2m}(k^{2}-q^{2})}\\
     &\times& \frac{1 }{ N}\sum_{l}\left<b_{l}^{2}e^{i({\bf q}
     -{\bf k}){\bf R}_{l}(t)} 
     e^{i({\bf k}-{\bf q}){\bf R}_{l}(0)}\right> \nonumber
\end{eqnarray}
and the other from multiple scattering
\begin{eqnarray}
    \label{eq:mulscattcorrection}
    v_{m}^{(2)} = & A &\int_{- \infty}^{+ \infty} dt 
    \int d^{3}{\bf q}e^{\frac{i\hbar t }{ 2m}(k^{2}-q^{2})}\\
    & & \mbox{} \times    \frac{1}{ N} \sum_{l,l^{'}, l\ne l^{'}}    
    \left[
     \left<b_{l}b_{l^{'}}    e^{i({\bf q}-{\bf k}){\bf R}_{l}(t)}    
     e^{i({\bf k}-{\bf q}){\bf
     R}_{l^{'}}(0)}\right> \right.\nonumber\\
    & & \mbox{}\mbox{} -  \left< b_{l}    
    e^{ i ( {\bf q}-{\bf k} ){\bf R}_{l}(t)} \right> 
    \left.\left<b_{l^{'}}    e^{i({\bf k}-{\bf q}){\bf R}_{l^{'}}(0)}
    \right> \right]\nonumber
\end{eqnarray}
and
\begin{equation}
    \label{eq:Multiplier}
    A=\frac{-i \rho }{ \left(2\pi\right)^{3} \hbar} 
    \left(\frac{2\pi \hbar^{2} }{m}\right)^{2}.
\end{equation}
In these equations $\rho$ is the number density of scatterers, $k$
and $q$ are the wave vectors of the incident and in-medium
neutrons (very close to identical for cold neutrons), $b_{l}$ is
the (spin dependent) scattering length operator of atom $l$, and
the averaging $\left<\right>$ is the usual trace over spins and
internal states of the scattering system. These expressions are
related to (but not identical with) the well-known dynamic
structure factors defined in the theory of thermal neutron
scattering\cite{Mar71} and make it clear that the second order
approximation to the optical potential is a function of the
dynamics and correlations of the scattering medium.

From this expression we can understand why the multiple scattering
correction to the optical potential is especially large for the
case of H and D gas molecules and much smaller for hydrogen atoms
embedded in the large polyatomic molecules that have been used for
the most precise n-p and n-d coherent scattering length
measurements in past work in liquid. First of all, the multiple
scattering term $v_{m}^{2}$ vanishes if the nuclei in the
polyatomic molecule possess uncorrelated nuclear spin directions
and uncorrelated relative motions, which is true to an excellent
approximation. From a physical point of view, multiple scattering
from uncorrelated nuclei cannot contribute to coherent scattering.
By contrast, as is well-known, the spins of hydrogen and deuterium
molecules are tightly correlated as a result of the identical
particle constraints, and the relative motions are perfectly
correlated for isolated molecules by the conservation of momentum.
Therefore the multiple scattering correction will be much larger
for H and D molecules than for larger polyatomic molecules.

In addition, one can see that the size of this effect will be
larger for hydrogen than for deuterium. In addition to the effect
of the larger mass, deuterium has a smaller ratio of incoherent to
coherent scattering, and as one can see from
Eq.~\ref{eq:hydrocorrection} the effect is proportional to this
ratio, which is anomalously large for hydrogen.

The expression for the dynamic structure factor in the corrections
calculated by Nowak can be evaluated exactly for hydrogen and
deuterium gas in thermal equilibrium in the rigid rotor
approximation.\cite{Mar71} The calculation draws on the work of
Young and Koppel\cite{You64} on the dynamic structure factor for
neutron scattering from hydrogen and deuterium molecules. The
rigid rotor approximation is known to be an excellent
approximation for slow neutron scattering in hydrogen and
deuterium. Since the neutron energy is too low to excite the first
vibrational level only transitions to rotational states need to be
taken into account.  These detailed calculations are presently in 
progress.

\subsection{Calculating the world scattering length average}
\label{sect:WorldAverageCalculation}
We consulted the existing compilations of previous
measurements\cite{Koe91,Rau00} for which we found the measurements of
the coherent scattering length of hydrogen
\cite{Cal75,Bur51,Squ55,Dic62,Koe71,Koe79,Kai79,Ham81} and
deuterium\cite{Dil71,Kai79,Ham81,Hur51,Shu51,Wor57,Gis63,Koe68,Nis74,Koe80}.
 We excluded all measurements that were not published in a refereed
journal and all measurements that were later retracted for
hydrogen\cite{Koe67,Koe75} and for
deuterium\cite{Nik55,Bar63,Bar68,Cop69,Gra79,Mei85} with the exception
of Ref.~\cite{Koe79}, which updates a previous value reported in
Ref.~\cite{Koe75} for isotopic purity.  Although the result of
Bartolini \etal\cite{Bar68} was never formally withdrawn it was
excluded here due to the fact that the result was $>10\sigma$ from the
world average.  The value quoted by Bartolini \etal\cite{Bar68} was
due to a reanalysis of the H$_{2}$O content of the sample used 4 years
prior in Ref.~\cite{Bar63}.  This large discrepancy was also discussed
by Dilg \etal \cite{Dil71}, and Nistler \cite{Nis74} who found this
result to be inconsistent.  In addition, there were two measurements
of the ratio of the bound coherent scattering lengths of hydrogen and
carbon whose originally reported uncertainties could be lowered by a
factor of 2 as a result of subsequent precision measurements of the
bound coherent scattering length of carbon, which is now known to be
\bcarbon \cite{Koe75}.  Including this new value for $b_{C}$ with the
measurement of Dickinson \etal\cite{Dic62} of
$b_{C}/b_{H}=(-1.775\pm0.004)$~fm gives $b_{H}=(-3.748\pm0.009)$~fm
for which permission was obtained from one of the coauthors
\cite{Pas02} to modify the original published result (original value:
$b_{H}=(-3.74\pm0.02)$~fm).  However permission was not obtained to
modify the measurement of Burgy \etal\cite{Bur51}.  Without permission
to modify this result we will only state the result here of how the
new value of $b_{C}$ affects the published value for $b_{H}$, but this
updated value will not be used in the calculation of the average value
of $b_{\mathit np}$.  From the measurement of Burgy \etal\cite{Bur51}
$b_{C}/b_{H}=(-1.756\pm0.005)$~fm gives $b_{H}=(-3.787\pm0.011)$~fm
(original value: $b_{H}=(-3.78\pm0.02)$~fm.  Note that Burgy \etal\
reported the ratio of the atomic scattering lengths as opposed to the
nuclear scattering lengths: we have converted their ratio into a ratio
of nuclear scattering lengths using the expression
$b_{C}/|b_{H}|_{nuc}=\left(b_{C}/|b_{H}|_{atom}\right)(1+b_{ne} /
|b_{H}|)+6b_{ne}/|b_{H}|$ where \bne\ is the neutron-electron
scattering length\cite{Koe86,Koe95,Kop95,Kop97}.

To calculate an average of all previous measurements we have followed
the following procedure given by Hagiwara \textit{et al} \cite{PDG02}. 
The data was combined into an average with results weighted inversely
with the size of their ($1\sigma$) uncertainties.  The result of this
evaluation is \bnpprevworld\ and \bndprevworld.  The reduced chi
square, $\chi^{2}_{r}$, of the average value for H is 3.6 and for D is
0.7.  Since the $\chi^{2}_{r}$ for H is greater than 1 we chose to
scale up the uncertainty by $\sqrt{\chi^{2}_{r}}$ based on the method
of Hagiwara \textit{et al} \cite{PDG02}.


\vfill

\begin{thebibliography}{00}
\bibitem{Fri83} J.L. Friar, B.F. Gibson, and G.L. Payne, Phys. Rev.
{\bf C28}  983(1983).
\bibitem{Ish99} S. Ishikawa, Phys. Rev. {\bf C59},  R1247 (1999).
\bibitem{Wit93} H. Witala, D. Huber, and W. Glockle, Few-Body Syst.
{\bf 14},  171 (1993).
\bibitem{Glo96} W. Glockle, H. Witala, D. Huber, H. Kamada, and J.
Golak, Phys. Rep. {\bf 274},  107 (1996).
\bibitem{Kie95a} A. Kievsky, M. Viviani, and S. Rosati, Phys. Rev.
{\bf C52}, R15 (1995).
\bibitem{Ros95} S. Rosati, M. Viviani, and A. Kievsky, Few-Body Syst.
Suppl. {\bf 8}, 21 (1995).
\bibitem{Kie95b} A. Kievsky, Few-Body Syst. Suppl. {\bf 9},
405 (1995).
\bibitem{Pud97} B. S. Pudliner \etal, Phys. Rev. {\bf C56},
1720 (1997).
\bibitem{Car94} J. Carlson and R. Schiavilla, Few-Body Syst.,
Suppl. {\bf 7}, 349 (1994).
\bibitem{Wir95} R. B. Wiringa, V. G. J. Stoks, and R. Schiavilla,
Phys. Rev. {\bf C51}, 38 (1995).
\bibitem{Kie93} A. Kievsky, M. Viviani, and S. Rosati, Nucl. Phys.
{\bf A551}, 241 (1994).
\bibitem{Bed97} P. F. Bedaque and U. van Kolck, Phys.Lett. {\bf B428}, 221
(1998).
\bibitem{Hal00} G. M. Hale, Los Alamos National Laboratory
(unpublished).
\bibitem{Bed98} P. F. Bedaque, H. W. Hammer, and U. van Kolck, Nucl.Phys. {\bf A646}, 444
(1999).
\bibitem{Ham99} H. W. Hammer, LANL e-Archives, nucl-th/9905036,
(1999).
\bibitem{Kol98} U. van Kolck, Nucl.Phys. {\bf A645}, 273 (1999).
\bibitem{Kap98} D. B Kaplan, M. J. Savage, and M. B. Wise, Phys.
Lett. {\bf B424},  390 (1998).
\bibitem{Geg98} J. Gegelia, LANL e-Archives, nucl-th/9802038 (1998).
\bibitem{Epe01} E. Epelbaum, H. Kamada, A. Nogga, H. Witala, W. Glockle,
and  U. G. Meissner, Phys. Rev. Lett. {\bf 86}, 4787 (2001).
\bibitem{Bla99} T. C. Black, H. J. Karwowski, E. J. Ludwig, A.
Kievsky, M. Viviani, and S. Rosati, Phys. Lett. B {\bf 471} (1999) 103.
\bibitem{Phi77} A. C. Phillips, Rep. Prog. Phys. {\bf 40}, 905 (1977).
\bibitem{Che91} C. R. Chen, G.L. Payne, J.L. Friar, and B.F. Gibson,
Phys. Rev. {\bf C44}, 50 (1991).
\bibitem{Dil71} W. Dilg, L. Koester, and W. Nistler, Phys. Lett. {\bf
B36}, 208 (1971).
\bibitem{Iof98} A. Ioffe, D.L. Jacobson, M. Arif, M. Vrana,
S.A. Werner, P. Fischer, G.L. Greene, and F. Mezei, Phys. Rev.
{\bf A58}, 1475 (1998).
\bibitem{Searsbook} V.F. Sears, \textit{Neutron Optics}, Oxford University
Press, 1989.
\bibitem{Ari94}  M. Arif, D. E. Brown, G. L. Greene, R. Clothier, and
K. Littrell, edited by C. G. Gordon, Proc. SPIE {\bf 2264}, 21
(1994).
\bibitem{Werbook} H. Rauch, S. Werner, \textit{Neutron Interferometry:
Lessons in Experimental Quantum Mechanics}, Oxford University
Press, 2000.
\bibitem{Moh00} P. Mohr and B.N. Taylor, Rev. of Mod. Phys.,
{\bf 72}, 351 (2000).
\bibitem{Mic59} A. Michels, W. de Graaff, T. Wassenaar, J.M.H.
Levelt, and P. Louwerse, Physica,'s Grav. {\bf 25}, 25 (1959).
\bibitem{Dym80} J. H. Dymond, \textit{The virial coefficients of pure gases
and mixtures: a critical compilation}, (New York, Oxford
University Press, 1980).
\bibitem{disclaimer} Certain trade names and company products are
mentioned in the text or identified in illustrations in order to
adequately specify the experimental procedure and equipment used.
In no case does such identification imply recommendation of
endorsement by the National Institute of Standards and Technology,
nor does it imply that the products are necessarily the best
available for the purpose.
\bibitem{Men00} d. Centreras, Mensor calibration report \# 13255 2000 
(unpublished).
\bibitem{Ichi92} M. Ichimura, Y. Sasajima, and M. Imabayashi,
Mat. Trans., JIM, {\bf 33},  449 (1992).
\bibitem{Comp93} A. Compaan, A. Wagoner, and A. Aydinli, Am. J.
Phys. {\bf 62},  639 (1994).
\bibitem{Liu02}C.-Yu Lin, Ph.D. Thesis, Princeton University
(2002).
\bibitem{Koe91} Koester, L., Rauch, H., and Seymann, E., Atomic Data
and Nuclear Data Tables 49, 65 (1991).
\bibitem{Stoup} John Stoup (NIST Test No. 821/265253-01).
\bibitem{CRC} In CRC Handbook of Chemistry and Physics, 68th Edition,
edited by R. C. Weast (Cleveland, Ohio, CRC Press, 1987).
\bibitem{Bas94} G. Basile, A. Bergamin, G. Cavagnero, G. Mana, E.
Vittone, and G. Zosi, Phys. Rev. Lett. {\bf 72}, 3133 (1994).
\bibitem{Sch91} J. Schmiedmayer, P. Riehs, J. A. Harvey, N. W. Hill,
Phys. Rev. Lett. {\bf 66}, 1015 (1991).
\bibitem{Now82b} E. Nowak, Z. Phys. {\bf B 49},  1 (1982).
\bibitem{Lip50b} B. A. Lippmann, Phys. Rev. {\bf 79}, 481 (1950).
\bibitem{Lip50a} B. A. Lippmann and J. Schwinger, Phys. Rev. {\bf 79},
469 (1950).
\bibitem{Lax52} M. Lax, Phys. Rev. {\bf 85}, 621 (1952).
\bibitem{Cal75} J. Callerame, D.J. Larson, S.J. Lipson, and R.
Wilson, Phys. Rev. {\bf C12}, 1428 (1975).
\bibitem{Sum84} D. Adli, and G. C. Summerfield, Phys. Rev. {\bf
A30}, 119 (1984).
\bibitem{Fri84} J.L. Friar, B.F. Gibson, G.L. Payne, and C. R. Chen,
Phys. Rev. {\bf C30},  1121(1984).
\bibitem{Ber86} G.H. Berthold and H. Zankel, Phys. Rev. {\bf C34}
1203 (1986).
\bibitem{Che89} C.R. Chen, G.L. Payne, J.L. Friar, and B.F. Gibson,
Phys. Rev. {\bf C39}, 1261 (1989).
\bibitem{Kie94} A. Kievsky, M. Viviani, and S. Rosati, Nucl. Phys.
{\bf A577}, 511 (1994).
\bibitem{Kie97b} A. Kievsky, Nucl. Phys. {\bf A624}, 125 (1997).
\bibitem{Kie97a} A. Kievsky, S. Rosati, M. Viviani, C.R. Brune, H.J.
Karwowski, E.J. Ludwig, and M.H. Wood, Phys. Lett. {\bf B406}, 292
(1997).
\bibitem{Epe02} E. Epelbaum, A. Nogga, W. Glockle, H. Kamada, U.
G. Meissner, and H. Witala, nucl-th/0208023v1, (2002).
\bibitem{Kol94} U. van Kolck, Phys. Rev. {\bf C49}, 2932 (1994).
\bibitem{Fri00} J. L. Friar, D. Huber, H. Witala, and G. L. Payne,
Acta Phys. Polon. {\bf B31}, 749 (2000).
\bibitem{Sto94} V. G. J. Stoks, R. A. M. Klomp, C. P. F. Terheggen,
and J. J. deSwart, Phys. Rev. {\bf C49}, 2950 (1994).
\bibitem{Fri93} J. L. Friar, G. L. Payne, V. G. J. Stoks, and J. J.
deSwart, Phys. Lett. {\bf B311}, 4 (1993).
\bibitem{Mac96} R. Machleidt, F. Sammarruca, and Y. Song, Phys.
Rev. {\bf C53}, 1483 (1996).
\bibitem{Rholt} R. Holt, private communication.
\bibitem{Zim02} O. Zimmer, G. Ehlers, B. Farago, H. Humblot, W.
Ketter, and R. Scherm, EJPdirect {\bf A1}, 1 (2002).
\bibitem{Kie98} A. Kievsky \etal, Phys. Rev. Lett. {\bf 81}, 1580
(1998).
\bibitem{Bre47a} G. Breit and P.R. Zilsel, Phys. Rev. {\bf 71}, 215 (1947).
\bibitem{Bre47b} G. Breit, Phys. Rev. {\bf 71}, 232 (1947).
\bibitem{Sum64} G. C. Summerfield, Annals of Physics: {\bf 26}, 72-80 (1964).
\bibitem{Die81} H. D. Dietze and E. Nowak, Z. Phys. {\bf B44}, 245
(1981).
\bibitem{Now82a} E. Nowak, Z. Phys {\bf B45}, 265 (1982).
\bibitem{Sea82} V. F. Sears, Phys. Rep. {\bf 82}, 1(1982).
\bibitem{Gol64} M. L. Goldberger and K. M. Watson, {\it Collision
Theory}, Wiley, New York (1964).
\bibitem{Mar71} W. Marshall and S. W. Lovesey, {\it Theory of
Thermal Neutron Scattering}, Oxford University Press (1971).
\bibitem{You64} J. A. Young and N. U. Koppel, Phys. Rev. {\bf A135},
603 (1964).
\bibitem{Rau00} H. Rauch and W. Waschkowski, in Landolt-Bornstein,
New Series I/16A, ed. H. Schopper, Chap. 6, Springer, Berlin
(2000).
\bibitem{Bur51} M.T. Burgy, G.R. Ringo and D.J. Hughes, Phys. Rev. {\bf 84}, 1160(1951).
\bibitem{Squ55} Squires, Proc. Roy. Soc. {\bf A230}, 19 (1955).
\bibitem{Dic62} W.C. Dickinson, L. Passell, and O. Halpern, Phys. Rev. {\bf 126}, 632 (1962).
\bibitem{Koe71} L. Koester and W. Nistler, Phys. Rev. Lett. {\bf 27}, 956 (1971).
\bibitem{Koe79} L. Koester, W. Nistler, Unpublished Munich (1979).
\bibitem{Kai79} H. Kaiser, H. Rauch, G. Badurek, W. Bauspiess, and U.
Bonse, Z. Physik {\bf A291},  231 (1979).
\bibitem{Ham81} S. Hammerschmied, H. Rauch, H. Clerc, U. Kischko, Z. Physik {\bf 302},  323 (1981).
\bibitem{Hur51} D.G. Hurst and N.Z. Alcock, Can. J. Phys. {\bf 29}, 36 (1951).
\bibitem{Shu51} C. G. Shull and E. O. Wollan, Phys. Rev. {\bf 81}, 527 (1951).
\bibitem{Wor57} J.E. Worsham, Jr., M.K. Wilkinson, and C.G. Shull, J. Phys. Chem. Sol. {\bf 3}, 303 (1957).
\bibitem{Gis63} W. Gissler, Z. Krist. {\bf 118}, 149 (1963).
\bibitem{Koe68} L. Koester, Z. Phys. {\bf 219},300(1968).
\bibitem{Nis74} W. Nistler, Z. Naturf. {\bf A29}, 1284(1974).
\bibitem{Koe80} L. Koester, Nuc. En. Ag. Nuc. Dat. Com. {\bf E-212U}, 64 (1980).
\bibitem{Koe67} L. Koester, Z. Phys. {\bf 198}, 187 (1967).
\bibitem{Koe75} L. Koester and W. Nistler, Z. Phys. {\bf A272}, 189 (1975).
\bibitem{Nik55} S.J. Nikitin, W.T. Smolyankin, W.Z. Kolganow, A.W. Lebedew, and G.S. Lomkazy, \textit{First International Conference on Peaceful Uses of Atomic Energy} (The United Nations, New York, 1956) {\bf 2}, 81 (1955).
\bibitem{Bar63} W. Bartolini, R. E. Donaldson, and L. Passell, Bull. Am. Phys. Soc. {\bf 8}, 477 (1963).
\bibitem{Bar68} W. Bartolini, R. E. Donaldson, and D. J. Groves, Phys. Rev. {\bf 174}, 313 (1968).
\bibitem{Cop69} P. Coppens, T.M. Sabine, A. Cryst. {\bf B25}, 2442 (1969).
\bibitem{Gra79} Granzer, Unpublished Viena (1979).
\bibitem{Mei85} J. Meier, diploma thesis TU-Munich, Reaktorstation 
Garching 1985.
\bibitem{Pas02} L. Passell, private communication.
\bibitem{Koe86} L. Koester, W. Waschkowski, and A. Kluever, Physica B 
\& C (Netherlands) {\bf B37}, 282 (1986).
\bibitem{Koe95} L. Koester, W. Waschkowski, L. V. Mitsyna, G. S.
Samsovat, P. Prokofjevs, and J. Tambergs, Phys. Rev. {\bf C51},
3363 (1995).
\bibitem{Kop95} S. Kopecky, P. Riehs, J. A. Harvey, and N. W. Hill,
Phys. Rev. Lett. {\bf 74}, 2427 (1995).
\bibitem{Kop97} S. Kopecky, J. A. Harvey, N. W. Hill, M. Krenn, M.
Pernicka, P. Riehs, and S. Steiner, Phys. Rev. {\bf C56}, 2229
(1997).
\bibitem{PDG02} K. Hagiwara {it et al.}, Phys. Rev. {\bf D66}, 010001-11 (2002).
\end{thebibliography}
\end{document}